\documentclass[lineno]{JFM-FLM_Au}
\usepackage{float}
\usepackage{caption}
\usepackage{subcaption}  
\usepackage[export]{adjustbox} 

\usepackage[normalem]{ulem}
\usepackage[percent]{overpic}
\usepackage{wasysym}

\usepackage{verbatim}
\newcommand{\HM}[1]{{\color{black}{#1}}}

\newcommand{\HMR}[1]{{\color{black}{#1}}}

\newcommand{\hOmega}{\hat{\Omega}}

\lefttitle{Nazari et al.}
\righttitle{Journal of Fluid Mechanics}

\title{Drag and Yielding of Rotating Bodies in Yield-Stress Fluids}

\author{Farshad Nazari\aff{1}, Akash Mittal\aff{2}, Kourosh Shoele\aff{2} \and Hadi Mohammadigoushki\aff{1,3}}

\affiliation{\aff{1}Department of Chemical and Biomedical Engineering, FAMU-FSU College of Engineering, Tallahassee, FL, 32310, USA
\aff{2}Department of Mechanical and Aerospace Engineering, FAMU-FSU College of Engineering, Tallahassee, FL, 32310, USA
\aff{3}National High Magnetic Field Laboratory, Florida State University, Tallahassee, FL, 32310, USA}

\corresau{Hadi Mohammadigoushki, \email{hadi.moham@eng.famu.fsu.edu}}

\begin{document}
\maketitle

\begin{abstract}
We investigate the settling dynamics of rotating objects in a yield-stress fluid by combining controlled experiments with numerical simulations. Experiments were conducted using cylinders and spheres of varying surface roughness, rotated within a Helmholtz coil and immersed in a Carbopol-based yield-stress fluid. Complementary numerical simulations employed a viscoplastic Herschel–Bulkley model to capture the coupled effects of sedimentation and rotation. To parameterize the problem, we define \HM{a dimensionless rotational velocity $\hat{\Omega}$} to characterize rotation and the \HM{Bingham number ($Bi$)} to characterize sedimentation. Measurements of the drag coefficient show a strong dependence on both surface roughness and rotation rate. \HM{The plastic drag coefficient is found to be inversely related to $\hat{\Omega}$ at fixed $Bi$, with rotation effectively reducing the resistance to motion.} Flow visualization reveals that enhanced rotation generates a plastic deformation zone in the orthogonal plane and promotes wall slip, while at \HM{low} \HM{$\hat{\Omega}$} a stagnation-point flow develops in the wake, gradually weakening and disappearing as \HM{$\hat{\Omega}$ increases}. In addition, the plastic drag coefficient decreases with increasing \HM{$Bi$} and approaches an asymptotic plateau at high \HM{$Bi$}. Numerical simulations reproduce the general scaling of drag with \HM{$\hat{\Omega}$} and \HM{$Bi$} but consistently underpredict experimental values, likely due to wall slip and nonlinear effects such as the stagnation-point flow not present in the model. \HM{The onset of sedimentation (yield limit) was also measured and found to increase with increasing rotation and to depend on surface roughness. Finally, simulations highlight scaling relations for drag coefficient, providing new insight into the interplay of sedimentation, rotation, and viscoplastic rheology.}
\end{abstract}



\section{Introduction}
The study of drag forces on objects moving through fluid media is one of the canonical problems in fluid mechanics~\citep{batchelor2000introduction}. Today, the study of low $Re$ drag coefficients is critical in fields such as biomedical engineering, nanotechnology, drilling muds, and slurry transport \citep{elgaddafi2012settling}. The focus of this study is on drag measurements on objects moving in yield stress fluids, a class of non-Newtonian fluids characterized by a finite stress threshold that must be exceeded before they flow. Yield stress fluids are commonly found in both natural and industrial settings, including in the food industry, biomedical applications, and drilling operations~\citep{balmforth2014yielding}.  \par 

{The settling motion of a single sphere  or cylinder  in yield stress fluids has been studied fairly well both in experiments ~\citep{ansley1967motion,jossic2001drag,merkak2006spheres,tabuteau2007drag,jossic2009drag,tokpavi2009experimental,holenberg2012particle,elgaddafi2012settling,ahonguio2014influence,murch2017growth} and numerical simulations~\citep{blackery1997creeping,tokpavi2008very,tokpavi2009experimental}. It has long been observed that when an object with a density different from the surrounding yield-stress fluid is introduced, a critical force threshold (yielding criterion) must be exceeded for the object to settle at a constant terminal velocity~\citep{ansley1967motion,jossic2001drag,merkak2006spheres,tabuteau2007drag,jossic2009drag,ahonguio2014influence}. This yielding criterion was suggested in early studies of Ansley and Smith~\citep{ansley1967motion}, and first determined analytically and numerically by Beris et al.~\citep{beris1985creeping} for a sphere settling in a Bingham fluid model. Later studies examined the yielding limit with numerical simulations~\citep{blackery1997creeping} and in experiments~\citep{jossic2001drag,tabuteau2007drag,jossic2009drag}. }\par

{Beyond yielding limit, studies have investigated the settling dynamics of spherical and cylindrical objects in yield stress fluids, and have assessed the drag coefficient~\citep{jossic2001drag,deglo2004sphere,mitsoulis2004creeping,chafe2005drag,tokpavi2009experimental,jossic2009drag,ouattara2018drag}, and details of the flow field around the settling object both in experiments~\citep{tokpavi2009experimental,holenberg2012particle,ahonguio2014influence,sgreva2020interaction} and numerical simulations~\citep{blackery1997creeping,mitsoulis2009simulation,chaparian2017cloaking,koblitz2018direct}. 
Experiments have mainly utilized polymeric gels, particularly those based on aqueous Carbopol solutions as model yield stress fluids~\citep{jossic2001drag,chafe2005drag,holenberg2012particle,holenberg2013ptv,fraggedakis2016yielding}. }\\ 

\begin{figure}[hthp]
\centering
\includegraphics[width=0.5\linewidth]{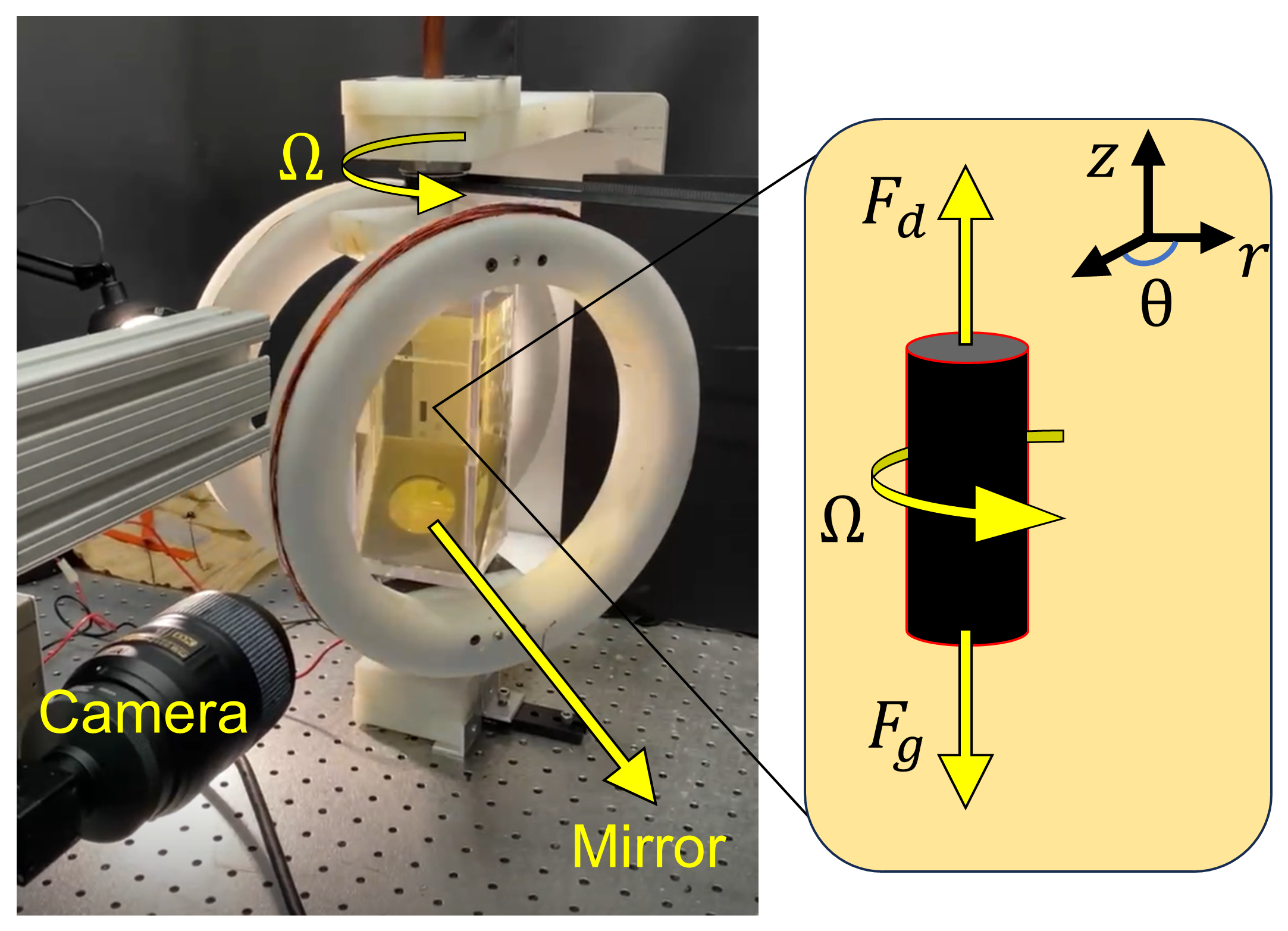}

\caption{A schematic of the rotational Helmholtz coil setup used to investigate the 
 sedimentation and rotational dynamics of cylindrical and spherical objects under a uniform magnetic field. Here $F_d$ and $F_g$ denote the drag and the gravitational forces, respectively.} 
\label{setup}
\end{figure}

{

While the flow of yield-stress fluids past stationary or translating particles is relatively well understood, scenarios involving the combined translational and rotational motion of a single object (e.g., a sphere or cylinder) in yield-stress fluids remain less explored~\citep{hewitt2018viscoplastic,balmforth2025implications}. \HM{In a small portion of their study, Hewitt and Balmforth investigated the axial motion with rotation of a cylinder in a Bingham fluid model, and showed in the limit of high and low Bingham numbers, drag force could be analytically estimated~\citep{hewitt2018viscoplastic}.} The combined rotation and sedimentation of objects in yield stress fluids have practical significance, particularly in biological contexts. For example, Helicobacter pylori (H. pylori), a gram-negative bacterium and leading cause of gastric ulcers, penetrates gastric mucus by flagellar propulsion\citep{li2023global, hooi2017global}. Experiments showed that under certain force conditions, H. pylori can become immobilized despite continuous flagellar and head rotation~\citep{celli2009helicobacter}. This result also hints at some critical yield limit for locomotion of H. pylori that is likely due to a balance between thrust generated by the tail and drag on the rotating head~\citep{nazari2023helical}. Therefore, the drag on the rotating head is particularly critical in shaping the bacterium’s locomotion, and understanding the yielding criterion, drag coefficient, and flow fields around rotating and translating objects is crucial for a potential therapeutic strategies involving H. pylori's infection. These fluid dynamics features remain unmeasured for bodies undergoing simultaneous translation and rotation in yield-stress fluids. \\

This study aims to systematically determine the yielding criterion, drag coefficient, and flow field characteristics around objects undergoing combined translational and rotational motion in a simple, non-thixotropic yield-stress fluid formulated with Carbopol-980. Two geometries, a sphere and a cylinder, with both smooth and roughened surfaces, are examined. Additionally, numerical simulations based on the Herschel-Bulkley fluid model are performed and compared with experimental results. \HM{This paper is organized as follows. Section~(\ref{methods}) describes the experimental methods. Section~(\ref{simulations}) presents the numerical model and simulation details. In Section~(\ref{dimensionless}), the relevant dimensionless numbers governing both the experiments and the numerical simulations are introduced and discussed. Section~(\ref{validation}) is devoted to the validation of the numerical model. In Section~(\ref{Results}), the experimental results are presented and compared with the numerical simulations. Finally, the main conclusions of the study and perspectives for future work are provided. }

\section{Experiments}
\label{methods}
\subsection{Material}

{The yield-stress fluid used in this study is prepared by mixing Carbopol-980 (sourced from Lubrizol) with triethanolamine. The concentration of Carbopol is maintained at 0.1 wt$\%$ for all experiments reported in this paper, with a fixed triethanolamine-to-Carbopol ratio of 1.5. Additionally, a Newtonian fluid based on corn syrup is prepared to provide a basis for comparison with the results obtained in the yield-stress fluid. The settling objects used in this experiment include a cylinder and a sphere, fabricated using a 3D printer (FormLab 3B). The sphere used had a diameter of \HM{$d=$} 8.07 mm, while the cylinder measured \HM{$d=$} 5 mm in diameter and \HM{$L=$} 14 mm in length. The dimensions of these objects are selected to have the same effective volume. For the roughened objects, the surface roughness was designed via 3D printing at an average of 250 $\mu$m (see Fig.~S1 of the supplementary materials). These objects are hollow, allowing for the insertion of additional weight. By adjusting the mass of the objects with weight, we could precisely control the gravitational force acting on them under specific experimental conditions. }\par 

\subsection{Experimental Setup}
Fluid rheology is characterized using an HR-10 TA rheometer coupled with a 40 mm cone and plate configuration, on which sandpaper was attached to the surface of the plate to avoid wall-slip. To measure yield criterion, the drag and analyze the flow field around settling and rotating objects, a custom-built Helmholtz coil setup, as shown in Figure~(\ref{setup}) is used. The Helmholtz coil system consists of two large circular coils connected to a DC power supply. When a DC current passes through the coils, it generates a uniform magnetic field in the central region. A motor applies a controlled angular velocity $\Omega$, causing the Helmholtz coils to rotate. Each object is embedded with a permanent magnet, and when placed in a chamber containing the fluid, it is rotated at an imposed rotational velocity by the rotating magnetic field of the Helmholtz coil. The settling speed is measured using particle tracking velocimetry (PTV). Additionally, the details of the flow field around the object are analyzed using a two-dimensional particle image velocimetry (PIV) method. For experiments related to PIV, the fluid was seeded with 50 ppm particles with an average size of 60-micron particles (from Potters Industries). The flow is illuminated in two planes (r-$\theta$), named as the orthogonal plane from now on, and the r-z plane, which is noted as the flow plane from now on.} Further details on experimental setup can be found in our previous publications~\citep{nazari2023helical,wu2022formation}. \par

\section{Numerical Simulations}
\label{simulations}

Here we consider an exterior flow problem of a steady settling  axisymmetric object with rotation in an infinite viscoplastic fluid given by the regularized Herschel-Bulkley model \citep{saramito2016complex, saramito2017progress}, written in the dimensional form as:
\begin{equation} \label{eq:viscoplastic1}
    -\nabla \cdot \boldsymbol{\sigma} + \nabla p = \boldsymbol{0}, \qquad
    \boldsymbol{\sigma} = 2 \eta(\dot{\gamma})\, \mathbf{D}(\mathbf{u}), \qquad
    \nabla \cdot \mathbf{u} = 0
\end{equation}

where \[\eta(\dot{\gamma}) = k \, \dot{\gamma}^{\,n-1} 
+ \sigma_y \, (\dot{\gamma}^2 + \epsilon^2)^{-1/2}
\] {is the effective viscosity which depends on the shear rate $\dot{\gamma}=|2\mathbb{D}|$.} 
The parameter $n > 0$ is the power-law index and $k$ is the associated consistency index. 
In addition, $p$ is the pressure, and $\boldsymbol{\sigma}$ and $\mathbf{D}$ are the deviatoric stress and rate-of-strain tensors, respectively. 
Moreover, $\mathbf{u}$ is the fluid velocity and $\epsilon$ is a small regularization parameter. 
The tensor norm is defined as $|\mathbf{A}|^2 = (\mathbf{A} : \mathbf{A})/2$ for any tensor $\delta \in \mathbb{R}^{3\times 3}$.

\HM{It is worth noting that the use of regularization methods for viscoplastic flows is known to involve inherent mathematical limitations, particularly regarding the convergence of the stress field and the accurate localization of rigid (unyielded) regions.  However, the model is shown by \citet{glowinski1981numerical} to exhibit monotonic convergence, with the velocity field approaching the exact solution in both the $H^1(\Omega)$ and $L^\infty(\Omega)$ norms, even when it results in a smooth transition between yielded and unyielded regions over a narrow band as opposed to sharply defined rigid zones. 
For problems with imposed kinematics, such as the swirl rotation considered here, viscoplastic fluid flow is dominated by strong boundary-induced shear near the particle surface. In such cases, regions where $D(\mathbf{u}_\epsilon) \to 0$  ($\mathbf{u}_\epsilon$ is the estimated flow velocity for an assumed regularized parameter ($\epsilon$) do not control the global dynamics, and the large quiescent zones typically observed in internal flows, such as Poiseuille or Couette configurations, are absent. Consequently, this model is found to be suited for the current study to examine velocity fields and flow topology, hydrodynamic drag, and the far-field flow structure as a function of the Bingham number.}

A  Galerkin Finite Element method (FEM) is used to numerically solve Eq. (\ref{eq:viscoplastic1}) \citep{eastham2020axisymmetric}. Its axisymmetric weak formulation at the $n^{\text{th}}$ iteration is:
\begin{subequations} \label{eq:FEM}
\begin{align}
\int_D2\eta\left(\dot{\gamma}^{n-1}\right)\left[\mathbb{D}(\mathbf{u}^n):\mathbb{D}(\mathbf{v}) + \frac{u_r^n v_r}{r^2}\right]&r\,dS - \int_D{p^n(\nabla\cdot \mathbf{v}) r}\,dS= 0, \\
\int_D q(\nabla\cdot \mathbf{u}^n) r \,dS &= 0,
\end{align}
\end{subequations}
\HM{where $D$ is the axisymmetric domain, $r$ is the radial distance,  $(\mathbf{u}^n, p^n)$ refers to unknown velocity and pressure fields to be solved at $n^{\text{th}}$ iteration while $\dot{\gamma}^{n-1}$ is calculated from the previous iteration's solution. Here $u_r$ denotes the radial component of the velocity vector. The test functions $(\mathbf{v}, q)$ are chosen to be Taylor-Hood Q$_2$--Q$_1$ shape functions. More details on the axisymmetric weak form for the Stokes equation and the validation of the model can be found in \citep{eastham2020axisymmetric,eastham2022squirmer}.} 

\begin{figure}[hthp]
\centering
\includegraphics[width=0.9\linewidth]{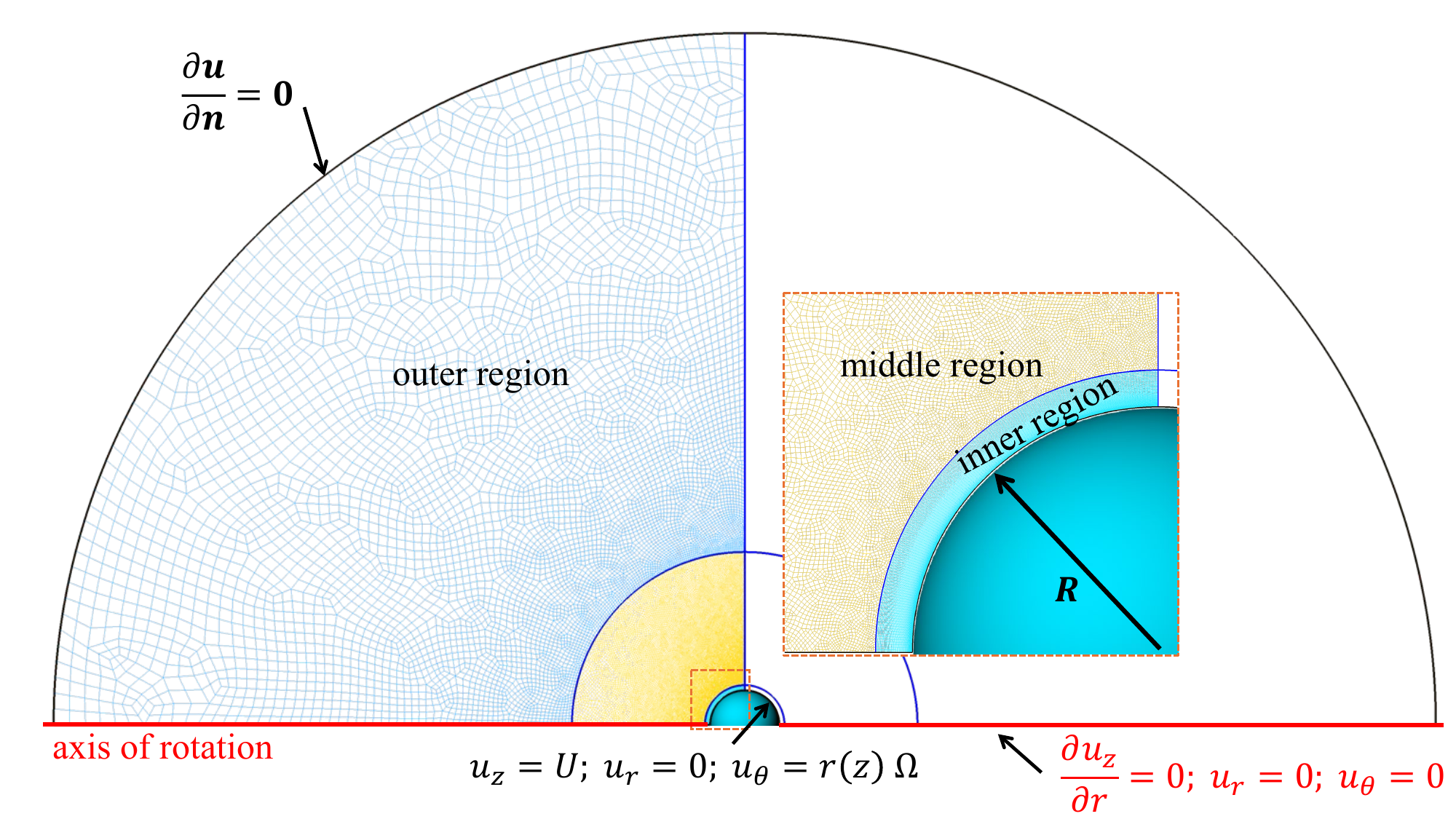}

\caption{Computational domain and imposed boundary conditions for the axisymmetric solver. The domain is partitioned into three concentric regions to optimize mesh density: an inner domain (thickness $0.15R$) for high-resolution capture of the yield surface, a middle region (thickness $5R$), and an outer region extending to $20R$ to minimize far-field effects. Here, $R$ is the radius of the sphere, while a similar setup is also used for the cylindrical case. Boundary conditions are defined as follows: at the object surface, a no-slip condition is applied where $u_z = U$ (sedimentation velocity), $u_r = 0$, and $u_{\theta} = r(z)\Omega$ (rotational velocity). Along the axis of rotation, symmetry conditions are enforced ($\partial u_z/\partial r = 0$). Here, $z$ and $r$ denote axial and radial coordinates, while $\theta$ represents the azimuthal angle.} 
\label{domain}
\end{figure}

\HMR{All viscoplastic simulations reported in this study were performed in a circular computational domain with an outer radius equal to 20 times the equivalent object radius, except in the model validation section, where a larger domain radius of 40$R$ was employed. The smallest Bingham number considered in the validation study was $Bi$ = 0.01, whereas all other simulations were conducted for $Bi\geq$ 0.1. Based on the theoretical predictions of \citet{taylor2025flow}, these domain sizes are sufficiently large to ensure that the yield surface remains fully contained within the computational domain.} The computational setup and boundary conditions, along with the mesh employed in this study, are shown in Figure~(\ref{domain}). Because the yield surface causes velocity field deformations to decay rapidly to negligible values, more elements can be concentrated near the object to better capture the yield surface. Additionally, the mesh is constructed to preserve discrete back–forward symmetry. Furthermore, to maintain consistency across all cases, simulations are performed for normalized equations by the particle radius and the maximum surface velocity, accounting for the combined effects of swirling rotation and sedimentation. 

\HM{We use a Picard iteration scheme to solve the nonlinear system Eq.~(\ref{eq:FEM}). The procedure starts with assuming a yield stress $\sigma_y$ and a relatively large regularization, and the values of both parameters are continuously updated in the solution procedure towards their final values. Each step with a given $Bi$ and $\epsilon$ also involves subiterations in which, at each iteration, the fluid Eq.~(\ref{eq:viscoplastic1}) is interpreted as a variable-viscosity Stokes equation with the current estimate of the effective constitutive relation. The system is iteratively solved until convergence with an absolute difference tolerance of $10^{-6}$. The procedure is repeated to determine the next step with a higher $\sigma_y{}$ and a smaller regularization parameter. Overall, 10 logarithmic increments of $Bi$ from the starting point of $Bi=0.1$ and 10 logarithmic increments of $\epsilon$ from the starting value of $\epsilon=0.1$ are considered. The overall solution procedure continues till the final convergence criteria of $10^{-6}$ is achieved.}

\HM{In classical perfect viscoplasticity models, the yield surface is defined by the condition $|\boldsymbol{\sigma}| = \sigma_y$, with the rigid region corresponding to $|\boldsymbol{\sigma}| < \sigma_y$, in accordance with the perfect viscoplasticity assumption. In the present work, however, we adopt a definition of the yield surface that is more consistent with a regularized formulation, identifying the rigid region based on yield term contribution in the stress as $\dot{\gamma}/\sqrt{\dot{\gamma}^2+\epsilon^2}\,=\,0.999$.}

\HM{The drag force and swirl torque are calculated from the pressure and deviatoric stress defined in Eq.~(\ref{eq:FEM}) as:
\begin{align} \label{eq:viscoplastic}
    F_d &=\int_S \quad\,\,\,\,\left[\left(-p\mathbf{I}+\tau (\dot{\gamma}) \right)\cdot \bm{n}\right] \cdot \bm{e}_z \, dS\\
    T_d &= \int_{S} \bm{r} \times [(-p\mathbf{I} + \mathbf{\tau}(\dot{\gamma})) \cdot \bm{n}] \cdot \bm{e}_z \, dS
\end{align}
where $S$ is the surface settling object and $\bm{r}$ represents the position vector from the axis of rotation to a point on the surface of the object. A systematic validation of the model is presented in section~(\ref{validation}) below. 
}

\section{Dimensionless Numbers}
\label{dimensionless}

\HM{When a particle translates at a constant terminal velocity in a yield-stress fluid, and inertial effects are negligible, the plastic drag coefficient, $C^*_d$ is used to quantify the resistance encountered by objects, and is defined as~\citep{jossic2009drag,jossic2001drag,mitsoulis2009simulation}:
\begin{equation}
  C^*_d = \frac{F_d}{A \, \sigma_y},
  \label{eq1}
\end{equation}
where $A$ is the projected area of the falling object and is given as $A = \pi d^2/4$. In addition, in numerical simulations, it is often convenient to normalize the drag force using the Stokes drag for a shear-thinning Herschel-Bulkley fluid. We therefore introduce Stokes' drag as:
\begin{equation}
  C_s = \frac{F_d}{6 \pi \eta U R},
\end{equation}
where $\eta$ is the viscosity defined as $k(U/R)^{n-1}$ with $U$ being the characteristic sedimentation velocity, and $R$ the radius of cross section of
the object. $k$ and $n$ are the consistency factor and shear-thinning index of the Herschel-Bulkley model, respectively. \HM{In addition, we define $C_s^V$ and $C_s^P$ for the contributions of pressure and shear stress in $F_d$ according to Eq. \ref{eq:viscoplastic}}. The dimensionless torque on a rotating cylinder is characterized using 
\begin{equation}
  T_s = \frac{T_d}{2\pi\eta U R^2},
\end{equation}
}

\HM{Another dimensionless group that appears in the literature is the gravitational yield number, defined as the ratio of the yield stress to buoyancy forces~\citep{jossic2009drag,jossic2001drag,mitsoulis2009simulation}:
\begin{equation}
    Y_G = \frac{\sigma_y}{g d_e\Delta \rho},
     \label{yg}
\end{equation}
where $g$ is the gravitational acceleration, $\Delta \rho$ is the density difference between the object and the fluid. The density of the particle is calculated as $\rho = m/V$, where $m$, and $V$ are particle mass and volume. Additionally, $d_e = (6V/\pi)^{1/3}$, and for the sphere $d_e = 2R$, while for the cylinder $d_e$ is the equivalent diameter of a sphere of the same volume as the cylinder. The dimensionless gravitational yield number has been widely used \HMR{(albeit in the absence of rotation)} to characterize the stability (or yield limit) of falling objects in yield‐stress materials. Specifically, the yield limit $Y_{G,\max}$ defines the threshold beyond which an object will not fall in a yield stress material (cessation of motion). }\par

\HM{For an object that undergoes both translation and rotation, two distinct viscous stresses arise: (i) a translational viscous stress associated with sedimentation ($U$), and (ii) a rotational viscous stress generated by the imposed angular velocity ($\Omega$). A meaningful characterization of the flow therefore requires two independent dimensionless groups to differentiate and quantify the relative contributions of translation and rotation. The appropriate choice of dimensionless numbers depends on which quantities are externally imposed and which are measured. In the experiments, the effective weight of the object $m$
(and hence the driving force for sedimentation) and the angular velocity 
$\Omega$ are prescribed, while the sedimentation velocity $U$ is measured. In contrast, in the numerical simulations both $U$ and $\Omega$ are directly imposed. Since our objective is to compare experiments and simulations, the dimensionless groups must be defined in a consistent manner across both approaches. To achieve this consistency, the experiments were designed such that $
U$ is effectively imposed in an indirect but controlled manner. In particular, for a given imposed rotational rate $\Omega$, the object’s effective weight was adjusted iteratively until the target sedimentation velocity was attained (see FIG.~S2 in the SI). The experimental tests are repeated at least three iterations to converge to the desired sedimentation velocity, wherein the results are obtained and reported. This procedure allows the experimental conditions to be mapped onto the same 
($U$,$\Omega$) parameter space as the simulations. To characterize the relative importance of translation, we therefore define a Bingham number that compares the material yield stress to the viscous stresses generated by translation as:
\begin{equation}
Bi = \frac{\sigma_y}{k (U/L_{c})^n},
\end{equation}
where the characteristic shear rate is $\dot{\gamma}= U/L_{c}$ and $L_{c}$ is the characteristic length (sphere radius or cylinder length). To quantify the relative importance of rotation, we nondimensionalize the rotational velocities using:
\begin{equation}
\hat{\Omega}  = \frac{ R\Omega}{U}.
\end{equation}
Large values of 
$\hat{\Omega}$ correspond to rotation-dominated flows, while small values indicate translation-dominated behavior.}





\HM{Finally, a Reynolds number can be defined for this study as: 
$Re = \rho U^{2-n}d^{n}/{k},$ where $\rho$ is the particle density. The Reynolds number assessed for this study is very small ($Re_\text{max} \approx 5.77\times10^{-3}$), and therefore, inertia does not affect the results. }

\section{Model validation}
\label{validation}
\begin{figure}[hthp]
\centering
\includegraphics[width=0.6\linewidth]{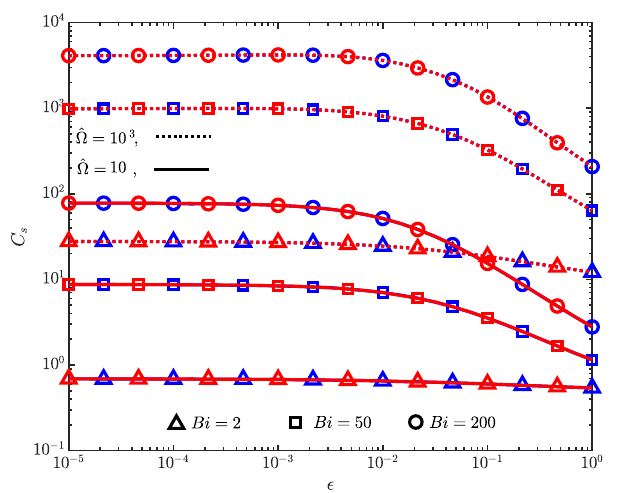}
\caption{Stokes' drag $C_s$ of a settling rotational sphere in a Herschel-Bulkely fluid model as a function of the regularization parameter $\epsilon$ for $Bi=2$ ($\triangle$), $50$ ($\square$) and $200$ ($\circ$) assuming different scaled rotational velocities of $\hat{\Omega} = 10, 10^{3}$. Two meshes are compared: The base mesh shown in Figure~(\ref{domain}) (mesh 1 with red symbols in this figure) and a more refined mesh with isotropic refinement and an average mesh size of $66\%$ of the base mesh shown with the blue color symbols. Here, only every fourth data point is shown with symbols.}
\label{regularization}
\end{figure}

\HM{\subsection{Regularization Parameter and Mesh Sensitivity}}
\HM{
The mesh resolution and the regularization parameter are systematically examined to ensure that the numerical model accurately captures the flow dynamics for the present configuration and across the intended parametric regimes. Three representative values of the Bingham number $Bi$ are selected for two distinct rotational velocities $\hat{\Omega}$ to assess the sensitivity of the solution (see Figure~(\ref{regularization})). In these tests, the settling particle is spherical and the surrounding fluid follows a Herschel--Bulkley rheology with $n=0.36$ (matched with experimental values), consistent with the parameters adopted throughout the remainder of this study.

The regularization parameter $\epsilon$ is progressively reduced from $1$ to $10^{-5}$, and the sensitivity of the Stokes' drag to the choice of $\epsilon$ is evaluated. For the base finite-element mesh (red curves in Figure~(\ref{regularization})), the results are found to converge for $\epsilon < 10^{-3}$.  The value of $\epsilon=5\times 10^{-4}$ is selected for all the numerical tests conducted in this work. A similar convergence behaviour is observed when the mesh is refined such that the average element size is reduced to $66\%$ of that of the base mesh, yielding nearly identical predictions. On the basis of this sensitivity analysis, the base mesh shown in Figure~(\ref{domain}) is adopted for all subsequent simulations.}

\HM{\subsection{ A rotating sphere in a Newtonian fluid }}
\HM{The first validation case considers a rotating sphere in a Newtonian fluid with $\hat{\Omega}=1$. In this regime, the swirl and translational components of the flow are decoupled, and the problem admits an analytical solution expressed in spherical coordinates $(r,\theta,\phi)$, representing the radial, polar and azimuthal directions, respectively, centred on a sphere of radius $R$ \citep{stokes1851effect}.}

\begin{equation}
\frac{u_r}{U} = -\cos \theta \left( 1 - \frac{3R}{2r} + \frac{R^3}{2r^3} \right); \hspace{0.5cm}
\frac{u_\theta}{U} = \sin \theta \left( 1 - \frac{3R}{4r} - \frac{R^3}{4r^3} \right); \hspace{0.5cm}
\frac{u_\phi}{U} = \frac{1}{\hat{\Omega}}\left(\frac{R}{r}\right)^3 \sin \theta,
\end{equation}

\begin{figure}[htbp]
\centering
\begin{minipage}[b]{0.49\linewidth} 
  \centering
  \begin{overpic}
  [width=0.86\linewidth]{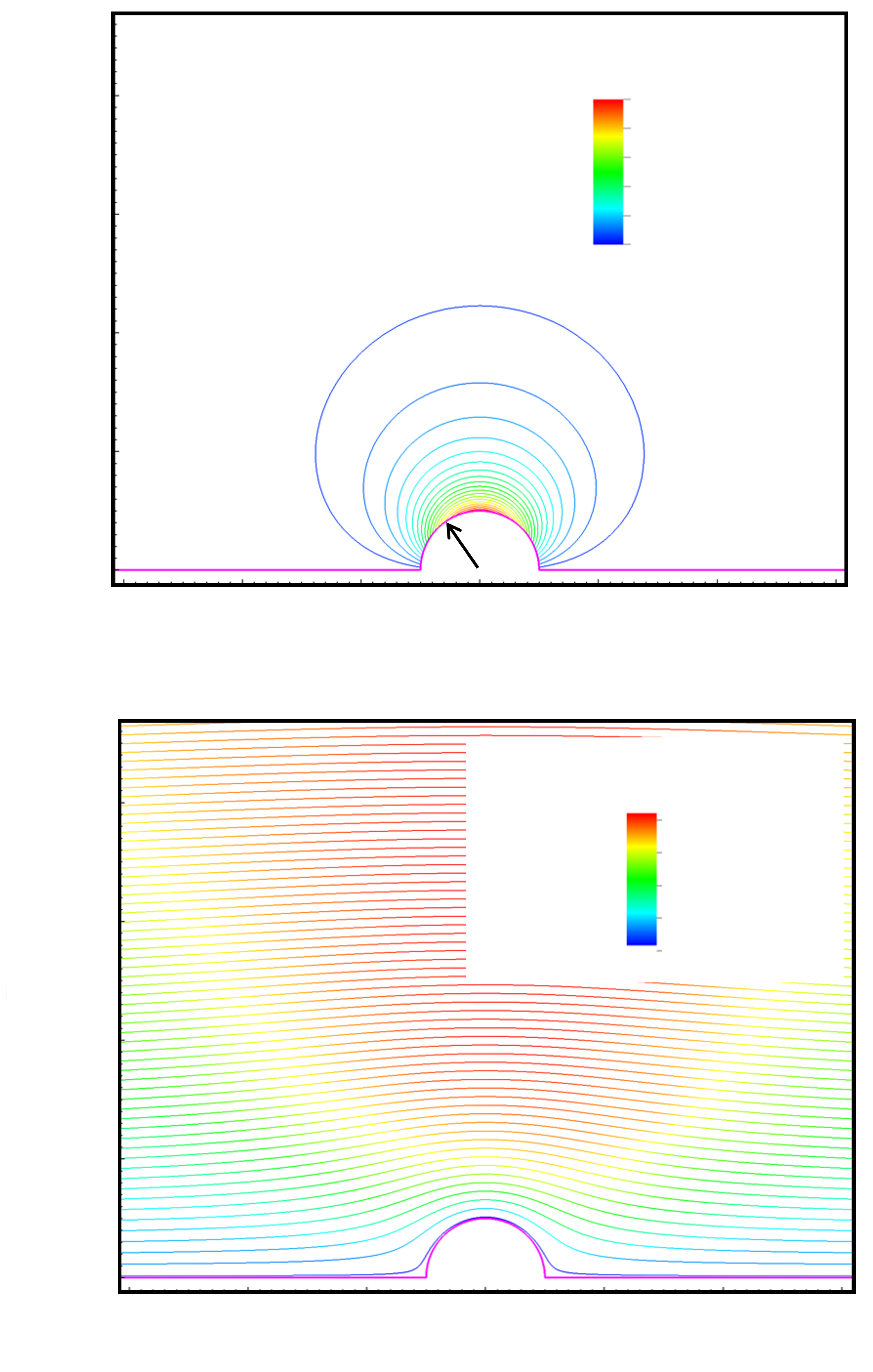}
  \put(45.5,96){\makebox(0,0){\fontsize{8pt}{9pt}\selectfont$\,u_{\theta}/{R\Omega}$}}
  
 \put(46,44){\makebox(0,0){\fontsize{8pt}{9pt}\selectfont$|\mathbf{u}-\mathbf{u}_{\text{analytical}}|/U$}}
 
  \put(4,77){\rotatebox{90}{\makebox(0,0){\fontsize{11pt}{12pt}\selectfont$r/R$}}}
    \put(4,26){\rotatebox{90}{\makebox(0,0){\fontsize{11pt}{12pt}\selectfont$r/R$}}}
  \put(35,52){\rotatebox{0}{\makebox(0,0){\fontsize{11pt}{12pt}\selectfont$z/R$}}}
    \put(35,0){\rotatebox{0}{\makebox(0,0){\fontsize{11pt}{12pt}\selectfont$z/R$}}}
 \put(9,56){\makebox(0,0){\fontsize{7pt}{8pt}\selectfont$-6$}}    
 \put(17.5,56){\makebox(0,0){\fontsize{7pt}{8pt}\selectfont$-4$}}    
 \put(27,56){\makebox(0,0){\fontsize{7pt}{8pt}\selectfont$-2$}}    
 \put(35,56){\makebox(0,0){\fontsize{7pt}{8pt}\selectfont$0$}}    
 \put(43.5,56){\makebox(0,0){\fontsize{7pt}{8pt}\selectfont$2$}}    
 \put(52,56){\makebox(0,0){\fontsize{7pt}{8pt}\selectfont$4$}}    
\put(61,56){\makebox(0,0){\fontsize{7pt}{8pt}\selectfont$6$}}  

 \put(9,4){\makebox(0,0){\fontsize{7pt}{8pt}\selectfont$-6$}}    
 \put(17.5,4){\makebox(0,0){\fontsize{7pt}{8pt}\selectfont$-4$}}    
 \put(27,4){\makebox(0,0){\fontsize{7pt}{8pt}\selectfont$-2$}}    
 \put(35,4){\makebox(0,0){\fontsize{7pt}{8pt}\selectfont$0$}}    
 \put(43.5,4){\makebox(0,0){\fontsize{7pt}{8pt}\selectfont$2$}}    
 \put(52,4){\makebox(0,0){\fontsize{7pt}{8pt}\selectfont$4$}}    
\put(61,4){\makebox(0,0){\fontsize{7pt}{8pt}\selectfont$6$}}  

 \put(7,7){\makebox(0,0){\fontsize{7pt}{8pt}\selectfont$0$}}    
 \put(7,16){\makebox(0,0){\fontsize{7pt}{8pt}\selectfont$2$}}    
 \put(7,24){\makebox(0,0){\fontsize{7pt}{8pt}\selectfont$4$}}    
\put(7,32.5){\makebox(0,0){\fontsize{7pt}{8pt}\selectfont$6$}} 
\put(7,41.5){\makebox(0,0){\fontsize{7pt}{8pt}\selectfont$8$}} 

 \put(7,59){\makebox(0,0){\fontsize{7pt}{8pt}\selectfont$0$}}    
 \put(7,67.5){\makebox(0,0){\fontsize{7pt}{8pt}\selectfont$2$}}    
 \put(7,76){\makebox(0,0){\fontsize{7pt}{8pt}\selectfont$4$}}    
\put(7,84.5){\makebox(0,0){\fontsize{7pt}{8pt}\selectfont$6$}} 
\put(7,93.15){\makebox(0,0){\fontsize{7pt}{8pt}\selectfont$8$}}

\put(47,82.2){\makebox(0,0){\fontsize{7pt}{8pt}\selectfont$0$}} 
\put(47,92.5){\makebox(0,0){\fontsize{7pt}{8pt}\selectfont$1$}} 

\put(49.,30){{\fontsize{6pt}{7pt}\selectfont$0$}} 
\put(49.,34.5){{\fontsize{6pt}{8pt}\selectfont$2\times10^{-4}$}} 
\put(49.,40.0){{\fontsize{6pt}{8pt}\selectfont$4\times10^{-4}$}} 

 \put(35.4,60){\makebox(0,0){\fontsize{6pt}{6pt}\selectfont$R$}}  

  \end{overpic}
  \put(-178,278){\makebox(0,0)[lt]{(a)}} 
  \caption*{}
\end{minipage}
\begin{minipage}[b]{0.49\linewidth}
  \centering
  \includegraphics[width=1\linewidth]{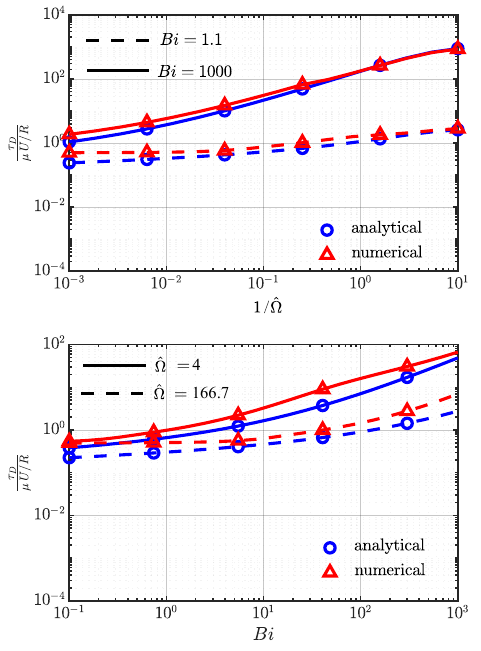}
    \put(-204,278){\makebox(0,0)[lt]{(b)}} 
  \caption*{}
\end{minipage}

\caption{(a) Comparison of flow for rotating sphere with $\hat{\Omega} =1$ in Newtonian fluid against analytical solution. The color shows the norm of the difference between the flow velocity vectors. (b) Comparison of shear stress on a long cylinder with tangential axial motion and swirl rotation against analytical results 
for a wide range of  $Bi$ and $\hat{\Omega}$ values. 
}
\label{FEMvalidation}
\end{figure}

\HM{The top panel of Figure \ref{FEMvalidation}(a) presents the contour plot of the swirl velocity component and the bottom panel of Figure \ref{FEMvalidation}(a) shows the norm of the difference between the analytical and computational velocity vectors. An excellent agreement is achieved between numerical results and those of analytical solution, with the numerical error remaining below $5 \times 10^{-4}$.}\\ 


\HM{\subsection{A rotating infinite cylinder in a Bingham fluid model.}}
\HM{The second validation is performed for a rotating and axially translating infinitely long cylinder in a Bingham viscoplastic fluid (with $n=1$). In this case, the rotation axis is perpendicular to direction of translation. This problem can be solved analytically as shown in \citep{hewitt2018viscoplastic}, resulting in analytical flow velocity that can be written in the cylindrical coordinate ($z, r, \phi$) as,}
\begin{equation}
\begin{aligned}
\frac{u_z}{U} &= \frac{r_p {Bi}}{C}\left[C^2 \log\left(\frac{r_p}{r}\right)-1+G(r)  \right], \frac{u_r}{U}= 0 \\
\frac{u_\phi}{U} &= \frac{r\,{Bi}}{2}
\left\{
S\left(\frac{r_p^2}{r^2}-1\right)
+\ln\!\left[
\frac{(1+S)\left(G(r)-S\right)}
{(1-S)\left(G(r)+S\right)}
\right]
\right\}
\end{aligned}
\end{equation}
where $G(r) \equiv \sqrt{C^2\left({r/r_p}\right)^2+S^2}$. 
Here, \(r_p\) denotes the radius of the yielded surface and \((C,S) = (\cos \Gamma, \sin \Gamma)\). The parameter \(\Gamma\), together with \(r_p\), is determined by satisfying the boundary conditions, which yields the tangential frictional shear on the cylinder as 
$\frac{\tau_D R}{\mu U} = r_p\, C\, {Bi}.$ \HM{This solution is based on the analytical results presented in \citep{hewitt2018viscoplastic}, with a necessary correction in the expression for \(u_\phi\) (Eq.~3.6 in \cite{hewitt2018viscoplastic}). For the computational model, the same mesh resolution used throughout the paper is employed here. The infinite geometry is approximated using a cylinder with an aspect ratio of $L/R=40$, and the reported frictional drag forces are calculated from the middle section of the cylinder to avoid end effects. Figure \ref{FEMvalidation}(b) shows the comparison between computational and analytical results of surface shear forces for a different range of $Bi$ and $\hat{\Omega}$ where a good match is found for a wide range of conditions. }

\begin{figure}[hthp]
\centering
\includegraphics[width=1.0\linewidth]{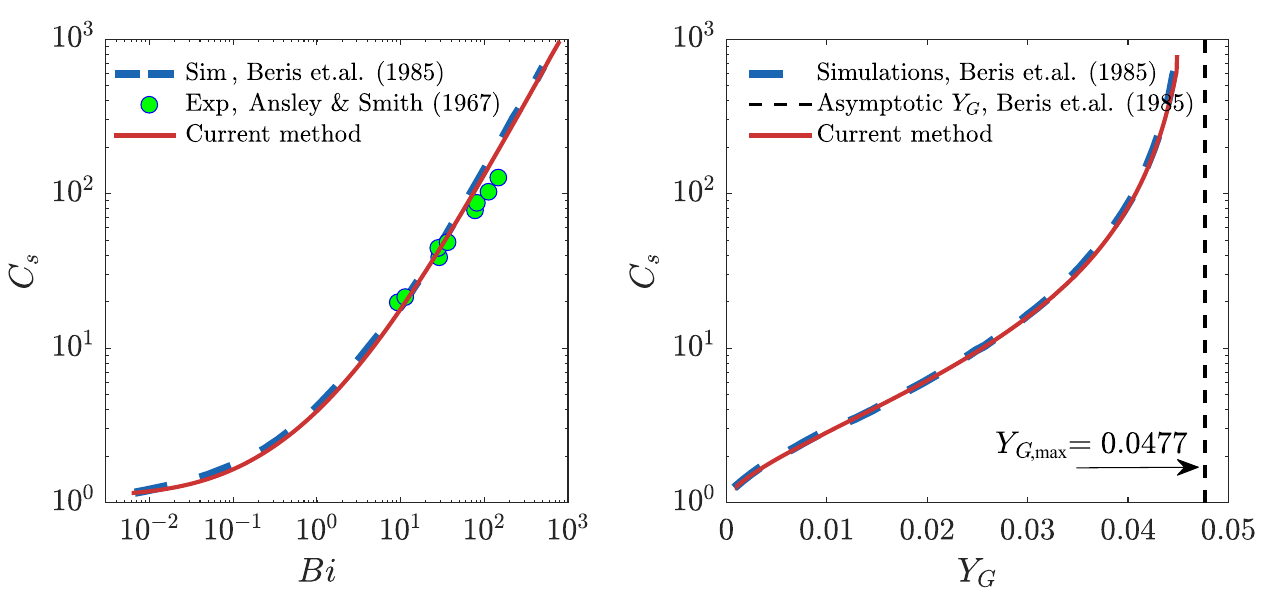}
\put(-390,176){\makebox(0,0)[lt]{(a)}} 
\put(-200,176){\makebox(0,0)[lt]{(b)}} 
\caption{\HM{Dimensionless Stokes drag as a function of $Bi$ number (a) and gravitational yield number (b). The simulations are compared with results of Beris et al.~\citep{beris1985creeping}, and experiments of Ansley \& Smith~\citep{ansley1967motion}.}}

\label{FEMvalidation2}
\end{figure}

\HM{\subsection{A non-rotating sphere in a Bingham fluid model}}
\HM{Finally, the computational technique is validated for a settling sphere in a Bingham fluid model with yield stress of $\sigma_y$ and viscosity of $\mu$. Here $\epsilon$ is selected based on a convergence study to be $0.0005$. \HMR{Figure~\ref{FEMvalidation2}(a)} compares the predicted Stokes drag on a sphere from the current method with previously reported results across different $Bi$ ranges, while \HMR{Figure~\ref{FEMvalidation2}(b)} shows a similar comparison across varying gravitational yield number defined in Eq.~(\ref{yg}). The simulations are carried out in a semicircular domain of $40R$, using a gradually coarsening mesh with element sizes ranging from $0.001R$ near the surface to $0.2R$ in the far field. The close agreement confirms the validity of the present computational approach.}

\section{Results and Discussion}
\label{Results}
\subsection{Fluid Characterization}
\begin{figure}[hthp]
\centering
\includegraphics[width=0.49\linewidth]{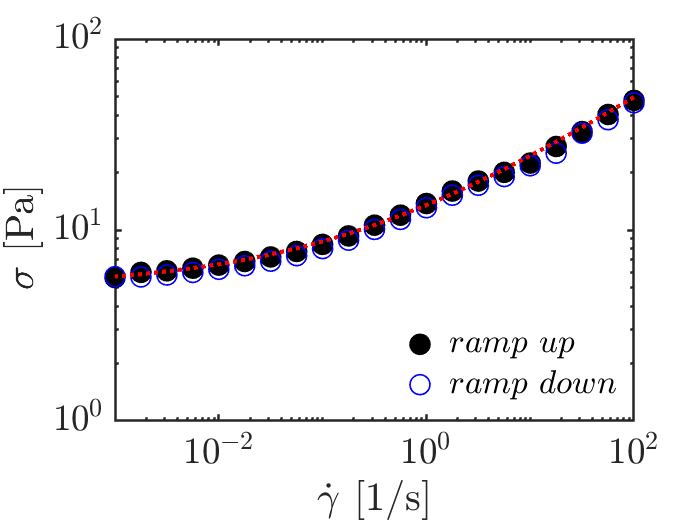}
\caption{Steady shear stress $\sigma$ as a function of shear rate $\dot{\gamma}$ for the yield stress fluid. The experimental data is fitted to the Hershel-Bulkley fluid model $\sigma = \sigma_y + k\dot{\gamma}^n$. Here $\sigma_y$ = 5.2 [Pa] is the yield stress, $k$ = 8.5 [Pa.s$^{n}$] is the consistency index, and $n$ = 0.36  is the shear-thinning index. 
}

\label{rheology}
\end{figure}

Figure~(\ref{rheology}) displays the steady-state flow curve of the yield-stress fluid measured over a broad range of applied shear rates. These flow curves were obtained by ramping shear rates up and down to assess potential thixotropic behavior in the fluid. As shown in Figure~(\ref{rheology}), the ramp-up and ramp-down curves overlap, indicating the absence of hysteresis and confirming that the fluid exhibits no measurable thixotropy. Furthermore, our measurements do not detect any first normal stress differences $(N_1)$ across the entire range of shear rates, suggesting that the fluid behaves as a simple viscoplastic, non-thixotropic material within the range of applied shear rates. The steady-state flow curve data are well-described by the Herschel-Bulkley fluid model (represented by the fitted red curves in Figure~(\ref{rheology}), with the corresponding fitting parameters provided in the figure caption. 

\subsection{Drag Coefficient}
In the following, we present the drag coefficient measurements and calculations as functions of rotation rate and sedimentation velocity.

\subsubsection{Effect of Rotation }
\begin{figure}[hthp]
\centering
\includegraphics[width=0.49\linewidth]{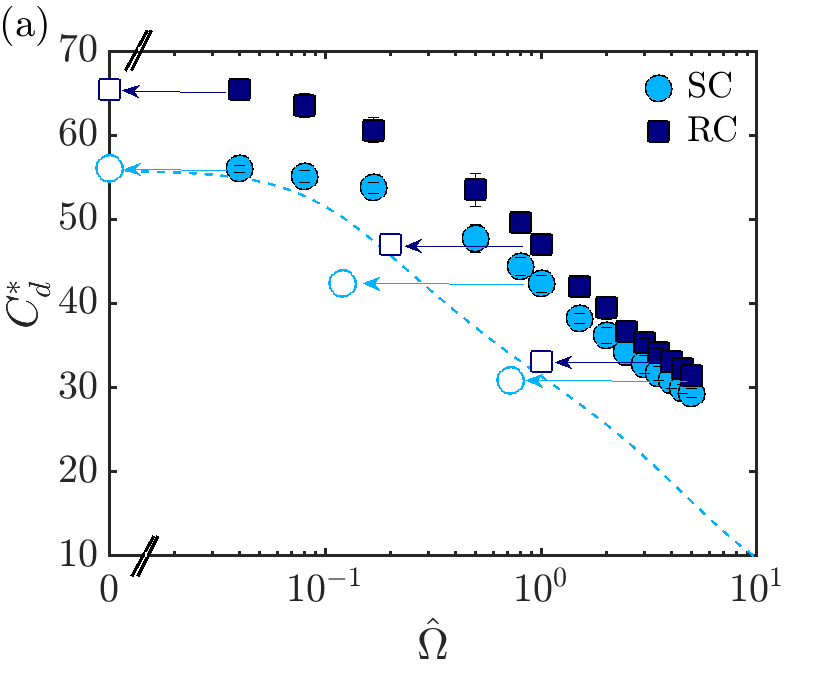}
\includegraphics[width=0.49\linewidth]{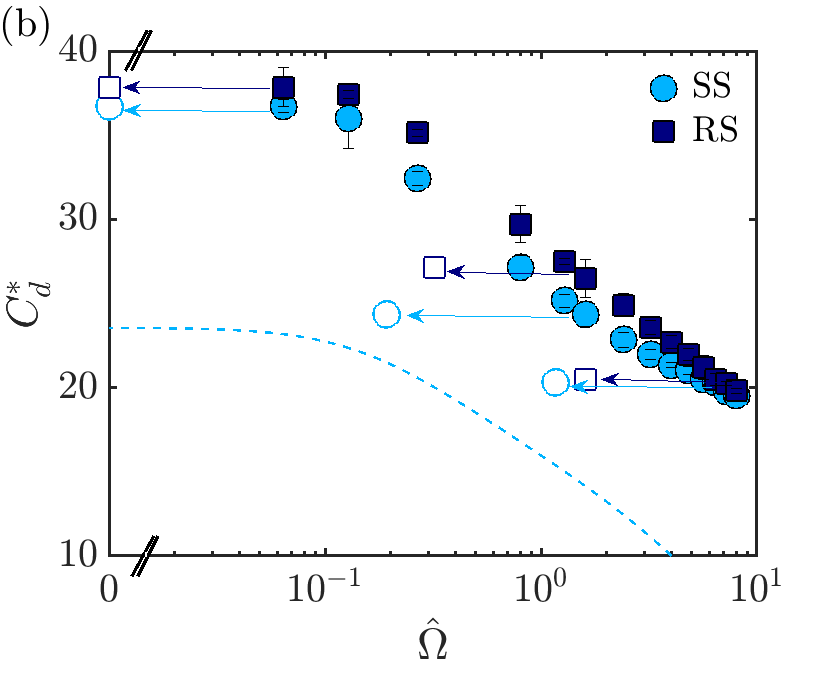}

\caption{Plastic drag coefficient as a function of dimensionless rotation rate \HM{$\hat{\Omega}$} for (a) cylinder and (b) sphere. The measured drag coefficients with a roughened surface are denoted as filled square and the smooth surface as filled circles. Smooth, roughened cylinders, smooth and roughed spheres are denoted as SC, RC, SS and RS, respectively. In addition, the hollow symbols are the effective \HM{$\hat{\Omega}_\mathrm{eff} $}  estimated based on the PIV data when effect of wall-slip is removed. The dashed curves show the results of numerical simulations. } 
\label{99}
\end{figure}

\HM{Figure~(\ref{99}) presents the measured plastic drag coefficient, $C_d^*$, for cylinders and spheres with varying surface roughness and $\hat {\Omega}$ numbers at a fixed Bingham number $Bi \approx 2.6$ for the cylinder and $Bi \approx 2.1$ for the sphere. Note that in the experiments reported in Figure~(\ref{99}), across all imposed angular velocities, $U$ is controlled and fixed (see the corresponding raw data for object's trajectory in FIG.~S2 of the SI).} Three clear trends emerge from these experiments. \HM{First, $C_d^*$ decreases monotonically with \HM{$\hat{\Omega}$} for all surface conditions and shapes. Second, roughened objects consistently exhibit higher plastic drag coefficients than smooth ones at a fixed rotation rate}. Finally, \HM{at high $\hat{\Omega}$ values (or high angular velocities)}, the difference in $C_d^*$ between smooth and roughened objects becomes negligible. 

\begin{figure}[hthp]
\centering

\includegraphics[width=0.65\linewidth]{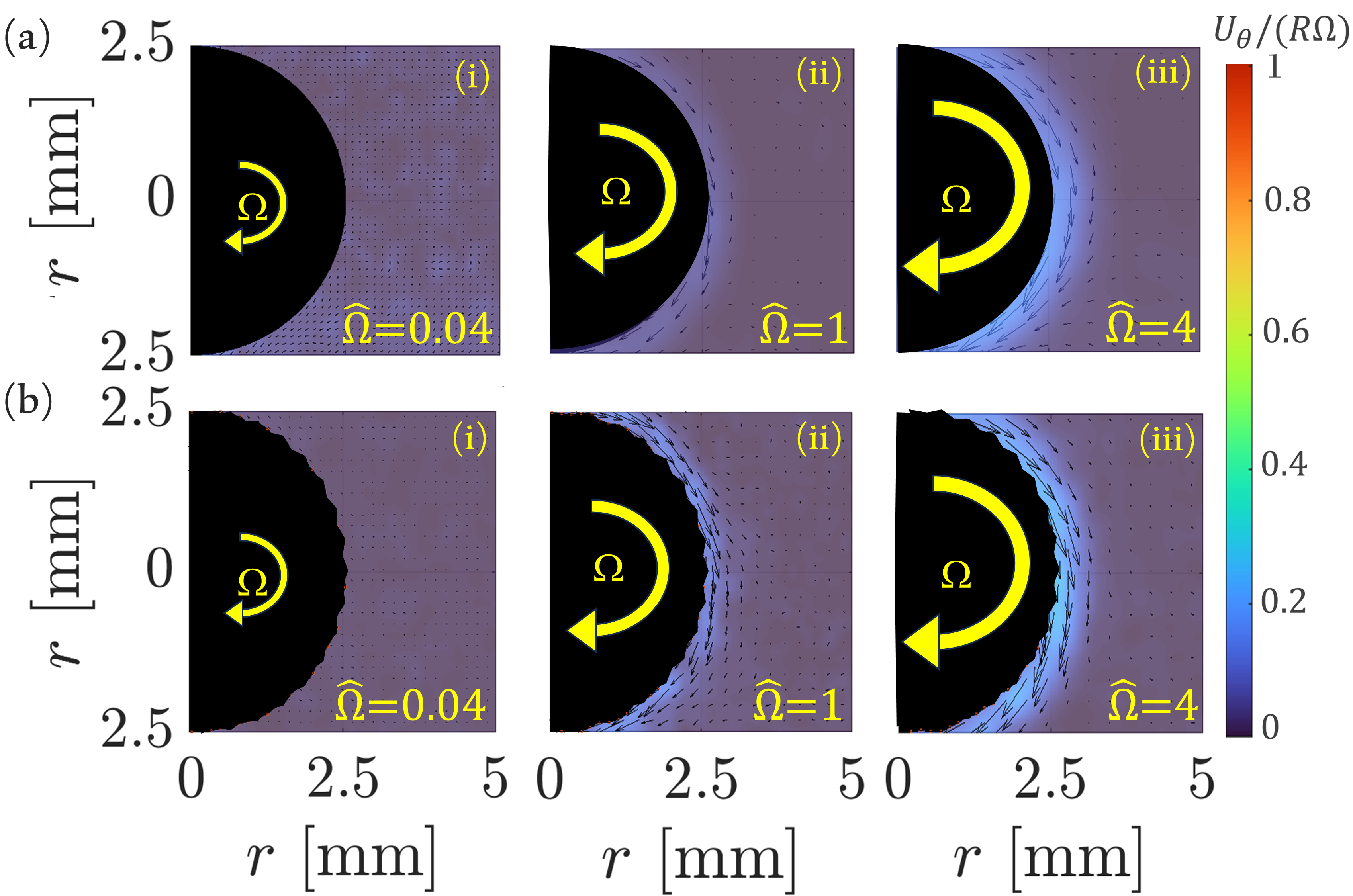}
\includegraphics[width=0.5\linewidth]{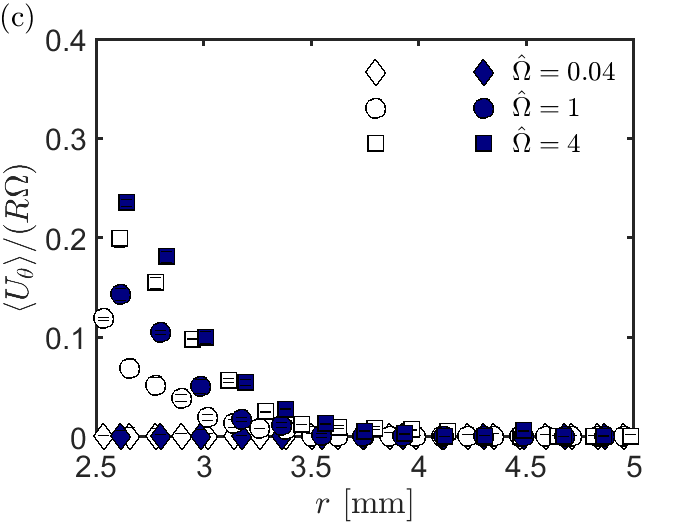}

\caption{2D averaged normalized velocity profile around the cylinder in orthogonal plane for smooth (a) and roughened (b) cylinders at a fixed \HM{$Bi \approx 2.6$} and various \HM{$\hat{\Omega}$ values (0.04, 1, and 4 from left to the right)}. (c) 1D normalized averaged rotational velocity measured for the yield stress \HMR{fluid} in the radial direction. Filled, and open symbols denote roughened and smooth surfaces, respectively. } 
\label{PIV_Y}
\end{figure}

To further elucidate the nature of the above trends, we performed particle image velocimetry (PIV) in a plane orthogonal to the cylinder's longest axis (or direction of sedimentation). Figure~(\ref{PIV_Y}) presents the two-dimensional, time-averaged, normalized fluid velocity fields ($\langle U_\theta \rangle/(R\Omega)$) around the cylinder at a fixed $Bi \approx2.6$ and varying rotation rates, for both smooth and roughened cylinders. At \HM{low $\hat{\Omega}$} numbers, despite a weak cylinder rotation, the surrounding medium shows negligible plastic deformation for both surface conditions. Two plausible scenarios may account for this observation. First, the yield-stress material adjacent to the cylinder may remain entirely unyielded, behaving as a rigid solid that permits slip of the cylinder. Alternatively, a yielded zone may exist but remain too small to be resolved within the spatial resolution of our measurements. Due to experimental constraints, we cannot conclusively distinguish between these two possibilities.

At \HM{ higher $\hat{\Omega}$ numbers ($\hat{\Omega} = 1$ and $4$)}, the plastic deformation zone shifts outward from the cylinder surface and becomes clearly resolvable in our experiments. Moreover, as \HM{$\hat{\Omega}$ increases}, this zone extends progressively farther into the surrounding material. These PIV results suggest that the variation in plastic drag coefficient with \HM{$\hat{\Omega}$ is directly related to the spatial extent of the plastic deformation zone surrounding the rotating body. As the rotation rate increases \HM{(or $\hat{\Omega} $ increases)}, a larger volume of material undergoes plastic deformation, which reduces the effective resistance of the surrounding medium. Consequently, a smaller drag force (or object mass) is required to maintain a fixed value of 
$Bi\approx2.6$.}  It is also evident from the PIV data that wall slip occurs in these experiments. We will come back to the effect of wall-slip later in the text.\\

Figure~(\ref{99}) also shows that the plastic drag coefficient for roughened surfaces is consistently higher than the measured values for the smooth surface. This result is consistent with the experiments of Jossic and Magnin, who also observed higher plastic drag coefficients for roughened cylinders relative to smooth ones at \HM{$\hat{\Omega} = 0$}~\citep{jossic2001drag,jossic2009drag}. PIV measurements in Figure~\ref{PIV_Y}(c) further indicate that, at a \HM{fixed $\hat{\Omega}$}, the extent of the yielded zone does not differ markedly between roughened and smooth surfaces. Instead, the key distinction lies in the degree of wall slip: it is reduced for roughened surfaces compared to smooth ones. These findings suggest that wall slip effectively reduces plastic drag in yield-stress fluids. \HMR{This result is consistent with predictions of Supekar et al. that showed for a Bingham plastic fluid, the wall-slip reduces the plastic drag coefficient of an infinitely long cylinder that sediments laterally~\citep{supekar2020translating}.} The plastic drag coefficient for the sphere exhibits trends similar to those observed for the cylinder, although the differences between roughened and smooth surfaces are somewhat less pronounced. In addition, the \HM{plastic} drag coefficient on spheres is, in general, smaller than that of cylinders. \HM{Corresponding PIV results for the sphere in the orthogonal plane are provided in FIG.~S3 of the SI.} \\

To further elucidate the above experimental trends, we performed numerical simulations using the Herschel-Bulkley viscoplastic fluid model with parameters matching the experimental conditions, including object dimensions, \HM{the rheological properties of the surrounding fluid, and imposed translational velocity}. The results, shown as dashed curves in Figure~(\ref{99}), generally capture the experimental trend of increasing plastic drag coefficient with \HM{$\hat{\Omega}$}. However, the simulations consistently underestimate the plastic drag coefficients for both the cylinder and the sphere. \HM{Previous numerical simulations with the Herschel-Bulkley viscoplastic fluid model also showed that the numerically predicted plastic drag coefficient $C^*_d$ underpredicts the experimental measurements for non-rotating cylinders in line with our results reported here~\citep{mitsoulis2009simulation,tokpavi2009experimental}.} \HM{Then, the natural question that arises is why simulations underpredict plastic drag coefficients? } Several factors may contribute to this discrepancy. One likely cause is the presence of wall slip in the experiments, which is absent in the simulations of this study. To evaluate the effect of wall-slip, we calculated an \HM{effective rotation rate $\hat{\Omega}_\mathrm{eff}$} for the experiments based on the rotation rate of the first fluid element adjacent to the object surface. For example, at an imposed \HM{$\hat{\Omega} = 1$, and roughened cylinder, the effective rotational velocity of the fluid is $\Omega_\mathrm{eff} \approx 0.2 \Omega$, and therefore, $\hat{\Omega}_\mathrm{eff}\approx 0.2$.} Here $\Omega$ denotes the imposed rotation of the solid body. This shifts the experimental data to the left in Figure~(\ref{99}) (depicted as empty symbols), thereby removing the possible influence of wall slip on the drag coefficient comparison. Interestingly, when the effect of wall slip is removed, the agreement between experiments and simulations improves substantially. At lower $\hat{\Omega}$ values, however, discrepancies remain, with simulations consistently under-predicting the experimentally measured drag coefficients for both spheres and the roughened cylinder. \HM{Another subtle consequence of the wall-slip in experiments is that in experiments where significant slip occurs, the yielded (plastic) region is observed to be thinner (or  closer to the object surface) than predicted numerically (see predictions of numerical simulations in FIG.~S4 in the SI).}\HMR{ As a result, the stress gradients in the vicinity of the rotating object can be significantly altered compared to simulations that enforce a no-slip boundary condition. This modification of the stress field can change the effective (force-velocity) relationship, which may lead to differences in the plastic drag coefficients.} \\

\begin{figure}[hthp]
\centering
\includegraphics[width=1 \linewidth]{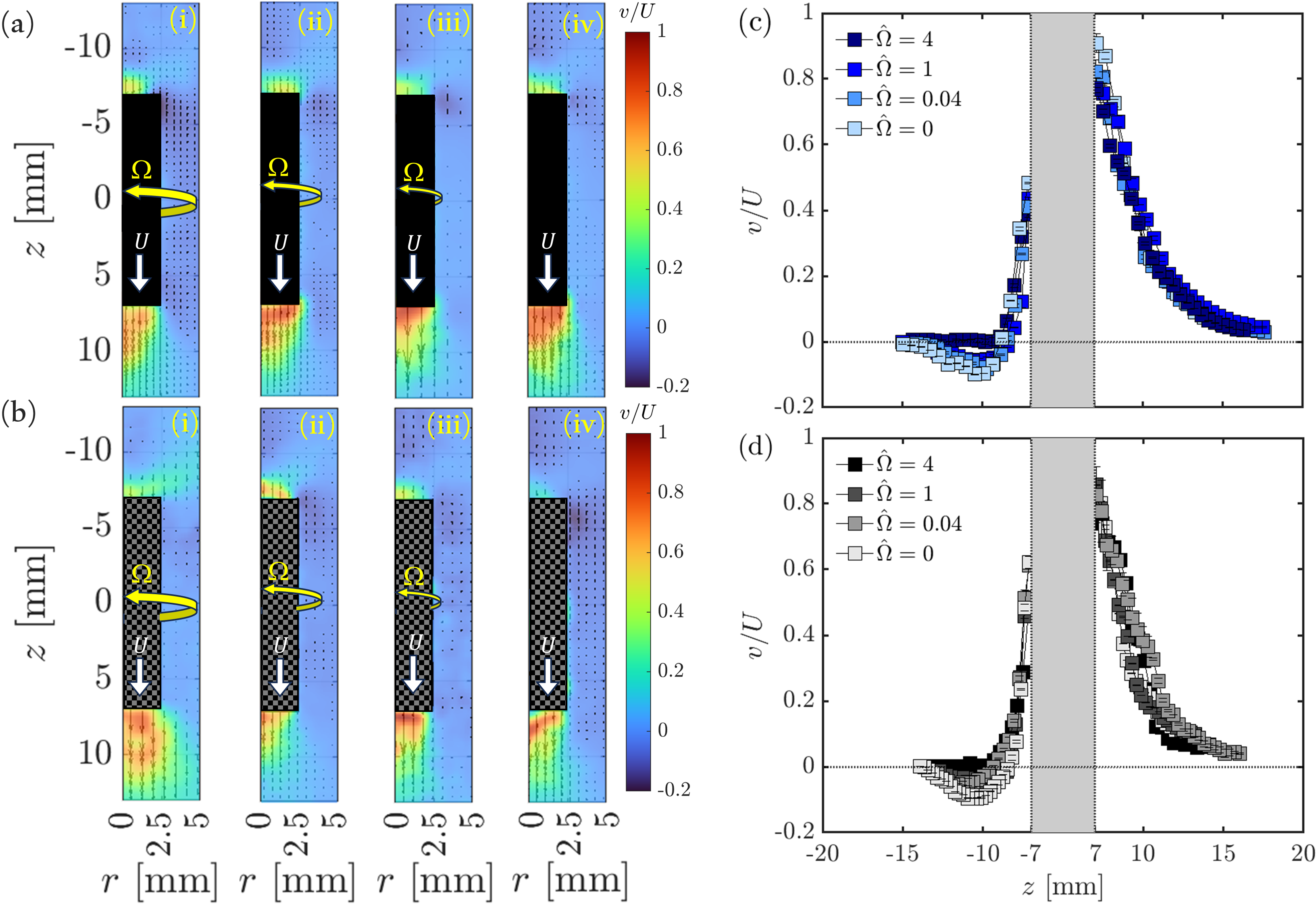}

\caption {2D and 1D averaged normalized velocity profile in flow plane ($v$ is fluid velocity component in z direction and $U$ is the sedimentation velocity of the smooth (a) and roughened (b) cylinder at a fixed \HM{$Bi\approx2.6$ and 
various $\hat{\Omega}$ values (4 (i), 1 (ii), 0.04 (iii), and 0 (iv)).} The averaged normalized velocity along the axis of the cylinder ($r=0$) for smooth (c) and roughened cylinder (d).}
\label{PIV-FP-cylinder}
\end{figure}

\HM{Beyond wall-slip, other factors may affect the drag in experiments in ways that are not captured by the viscoplastic simulations. } Figure~\ref{PIV-FP-cylinder}(a,b) show the averaged 2D velocity vectors and velocity magnitude around the falling cylinder. In the absence of any rotation \HM{($\hat{\Omega} = 0$)}, the velocity field is asymmetric around the cylinder with the upstream flow clearly showing a stagnation point flow accompanied by a negative wake (see also the averaged velocity at the center-line of the cylinder in Figure~\ref{PIV-FP-cylinder}(c)). As the rotational velocity increases, the stagnation point flow becomes less significant and eventually disappears at the highest \HM{$\hat{\Omega}$} number. Changing the surface roughness does not significantly affect the details of the flow field around the cylinder at low \HM{$\hat\Omega$} numbers. A similar trend is observed for flow past the smooth and roughened spheres, as shown in FIG.~S5 in the SI. In contrast, the counterpart results in the Newtonian fluid (corn syrup) show a symmetric flow field, highlighting how the non-Newtonian behavior leads to asymmetries in the flow profiles (see FIG.~S6-S7 in the SI).\\

It is worth noting that numerical simulations using the Herschel-Bulkley viscoplastic fluid model do not capture the flow asymmetry observed experimentally around the falling object. Interestingly, as the negative wake feature is dampened in the experiments at higher \HM{$\hat\Omega$ numbers}, the discrepancy between drag coefficients of simulations and experiments diminishes (see empty symbols corrected for the effect of wall-slip in Figure~(\ref{99})). This trend suggests a strong correlation between the formation of flow asymmetry (particularly the negative wake) around the falling object and the resulting drag coefficient. A similar asymmetry in the fore-and-aft direction has been reported for a sphere settling in yield stress fluids based on carbopol~\citep{putz2008settling,tokpavi2009experimental,sgreva2020interaction,holenberg2012particle}. This flow phenomenon has been suggested to be associated with either the viscoelasticity nature of the yield stress fluid or presence of aging or thixotropy~\citep{putz2008settling,tokpavi2009experimental,sgreva2020interaction,holenberg2012particle}. Recent finite element simulations using Elasto-Viscoplastic (EVP) fluid models predicted the presence of for-aft-asymmetry around a falling sphere and showed that yield stress significantly enhances the negative wake behind the objects in an EVP fluid model \citep{yazdi2023sedimentation,abbasi2024sedimentation,fraggedakis2016yielding}. Following preparation steps noted in ~\cite{dinkgreve2018carbopol}, the flow curves of the yield stress fluid were characterized both via shear rate ramp up followed by a shear rate ramp down, and showed no sign of thixotropy or hysteresis. \HM{In addition, the non-linear rheological measurements of the yield stress fluid used in this study do not show any signs of normal stress differences or viscoelasticity, which suggests that the plastically deformed fluid around the object is unlikely to form normal stresses associated with viscoelasticity.} These rheological characterizations further support the expectation that no fore-aft asymmetry should arise in the flow field for the fluid used in this study. Nevertheless, the origin of the reported negative wake around the object in this work remains unclear. \HM{Despite this, prior numerical simulations based on EVP fluid model have suggested that, at fixed terminal velocity or Stokes' drag, the plastic contribution to the drag decreases with increasing viscoelasticity of the EVP fluid model~\citep{fraggedakis2016yielding}. In this study an EVP fluid model based on Oldroyd type formulations was used, and it is possible that drag reduction was a consequence of that model choice. It is worth noting that experimental evidence isolating the role of elasticity on drag in yield stress fluids, particularly in the presence of imposed rotation is currently lacking. This is the subject of our future work.}\\

\begin{figure}
    \centering
  \vspace{0.5cm}
    \begin{subfigure}[t]{0.48\textwidth}
        \includegraphics[width=\linewidth]{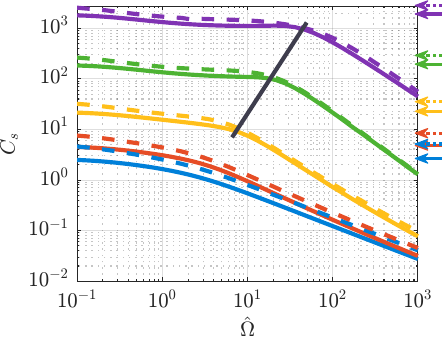}
        \put(-180,155){\makebox(0,0)[lt]{(a)}} 
    \end{subfigure}
    \hfill
    \begin{subfigure}[t]{0.47\textwidth}
        \includegraphics[width=\linewidth]{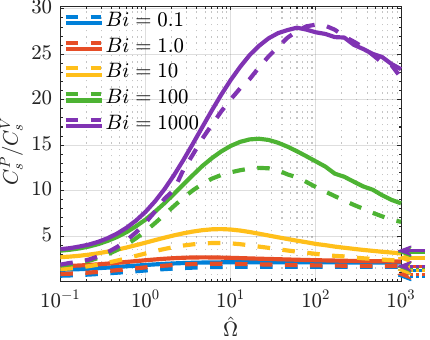}
        \put(-180,155){\makebox(0,0)[lt]{{(b)}}}
    \end{subfigure}

    \vspace{0.5em} 

    \begin{subfigure}[t]{0.48\textwidth}
        \includegraphics[width=\linewidth]{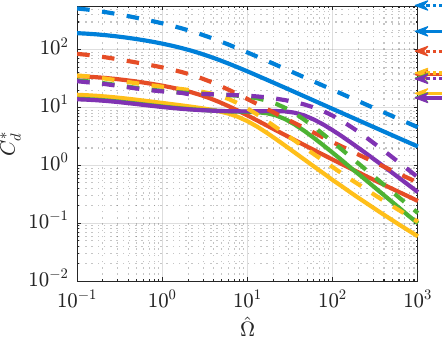}
        \put(-180,155){\makebox(0,0)[lt]{{(c)}}}
    \end{subfigure}
    \hfill
    \begin{subfigure}[t]{0.47\textwidth}
        \includegraphics[width=\linewidth]{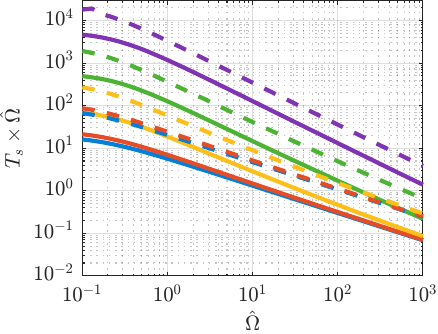}
        \put(-180,155){\makebox(0,0)[lt]{{(d)}}}
    \end{subfigure}

    \caption{Simulation results for (a) normalized drag forces $C_s$, (b) ratio of the pressure drag to shear-induced drag, (c) $C^*_d$ and (d) normalized torque as a function of $Bi$ numbers for different $\hat{\Omega}$. The continuous and dashed curves correspond to simulations on the sphere and the cylinder, respectively. The arrows show the results for non-rotating objects at different $Bi$ values.}
    \label{fig:FdversusRo_COMP}
\end{figure}

\HM{
To gain deeper insights into the dependence of the drag coefficients on the rotation, we carried out a systematic series of numerical simulations. Figure~(\ref{fig:FdversusRo_COMP}) presents the variation of drag force across a wide range of rotational conditions represented by $\hat{\Omega}$. The line colors indicate different values of $Bi$, with solid lines corresponding to the sphere and dashed lines to the cylinder. 
Two drag coefficients $C_s$, and $C^*_d$ are applied to illustrate the changes in drag force in Figure~\ref{fig:FdversusRo_COMP}(a and c). Additionally, Figure~\ref{fig:FdversusRo_COMP}(b) isolates the contribution of pressure versus viscous frictional forces, while Figure~\ref{fig:FdversusRo_COMP}(d) displays the induced torque on the rotating object. For each curve, the corresponding results for non-rotating objects are indicated with arrows on the rightmost side of the figures.}\par 

\HM{The most pronounced variations in drag force occur at large $\hat{\Omega}$ numbers, where azimuthal rotation controls the formation and shape of the yielded flow region around the rotating object.} \HMR{In this regime, both $C^*_d$ and $C_s$ exhibit a clear power-law dependence on $\hat{\Omega}$, with a slope that strongly depends on $Bi$. At smaller $\hat{\Omega}$, the scaling approaches an approximately constant value of $C_s \sim \hat{\Omega}^{-0.1}$, except for cases with thick yield regions associated with $Bi<1$.}\HM{ Therefore, for rotational rates small relative to the sedimentation velocity, the drag coefficient is a weak function of $\hat{\Omega}$, consistent with the limited experimental observations reported in Figure~(\ref{99}). \HMR{A notable feature is the transition to a stronger power-law dependency} for all cases as shown in Figure~\ref{fig:FdversusRo_COMP}(a). The transition between the two regimes shifts to larger $\hat{\Omega}$ as $Bi$ increases. The intersection points of the \HMR{transition} at small and large $\hat{\Omega}$ are well approximated by a curve with slope $1.64$, and is shown by the black line in Figure~\ref{fig:FdversusRo_COMP}(a). } 

\HM{Figure~\ref{fig:FdversusRo_COMP}(b) presents the ratio of the pressure contribution \HMR{($C^P_s$)} to the viscous contribution \HMR{($C^V_s$)} in the the drag force, i.e.\ $\int_S -p\mathbf{n}\cdot\mathbf{e}_z\,\mathrm{d}S$ relative to $\int_S (\boldsymbol{\tau}\cdot\mathbf{n})\cdot\mathbf{e}_z\,\mathrm{d}S$, as a function of $\hat{\Omega}$. The pressure-to-viscous drag ratio reaches a maximum near the intersection of the two $C_s$ power-law regimes discussed above, attaining values of approximately $28$ for $Bi=1000$. The variation of this ratio with $\hat{\Omega}$ is closely linked to the shape of the yielded region surrounding the settling body and will be discussed below. }
\HMR{This observation is also aligned with the scaling argument developed in Appendix~A, where we discuss a physical interpretation
of this trend in the high-swirl limit. When the surface traction becomes
primarily azimuthal, only its meridional projection contributes to axial drag,
giving
\begin{equation}
C^{*}_{d,\tau}
\sim
C_{\tau}\left(1+\hat{\Omega}^{2}\right)^{-1/2}.
\end{equation}
A confined yielded layer can also generate a pressure contribution,
\begin{equation}
C^{*}_{d,p}
\sim
C_p Bi^{2/(n+1)}
\hat{\Omega}^{-(3n+1)/(n+1)},
\end{equation}
which reduces to $C_p Bi/\hat{\Omega}^{2}$ for a Bingham material. Thus, the
pressure-to-viscous drag ratio in Fig.~10(b) is controlled not only by the
magnitude of rotation, but also by how rotation reshapes and confines the
yielded region.}

\HM{Our simulations suggest that scalings of plastic drag coefficient $C^*_d$ (as shown in Figure \ref{fig:FdversusRo_COMP}(c)) with respect to $\hat{\Omega}$ follow similar trends to those reported for $C_s$. In the small $\hat{\Omega}$ regime, the plastic drag coefficient is only marginally affected by the object rotating condition, and results effectively approach those of non-rotating objects below $\hat{\Omega} < 0.1$. The experimental results of Figure~(\ref{99}) that are performed at a fixed Bingham number of $Bi\approx2.6$ and $0< \hat{\Omega} < 5$, suggest that for the cylinder $C_d^* \sim \hat{\Omega}^{-0.24}$ for the roughened cylinder and $C_d^* \sim \hat{\Omega}^{-0.22}$ for the smooth cylinder over a range of $0.5\leq  \hat{\Omega} \leq 5$. In addition, $C_d^* \sim \hat{\Omega}^{0.18}$ for roughened sphere and $C_d^* \sim \hat{\Omega}^{0.14}$ for smooth sphere in the range of $ 0.2\leq \hat{\Omega} \leq 8$.} These scalings are within the range predicted by the numerical simulations. \\

\HM{Finally, Figure~\ref{fig:FdversusRo_COMP}(d) shows the variation of the imposed torque with $\hOmega$ through the quantity $T_s/\hOmega$. Similar trends are observed for both the sphere and the cylinder. In the rotation-dominated regime, the torque decreases with $\hOmega$ following a power-law behaviour for all tested $Bi$, where the curves become steeper with increasing $Bi$. This approximately linear power-law trend starts from $\hOmega\ge1$, after which the torque/rotation coefficient continues to decrease. The deviation from power law occurs at low rotational rates where the translational motion is the primary factor that controls the yielded region near the object. }

\begin{figure}[htbp]
  \centering
    \begin{overpic}[width=1\linewidth,tics=3]{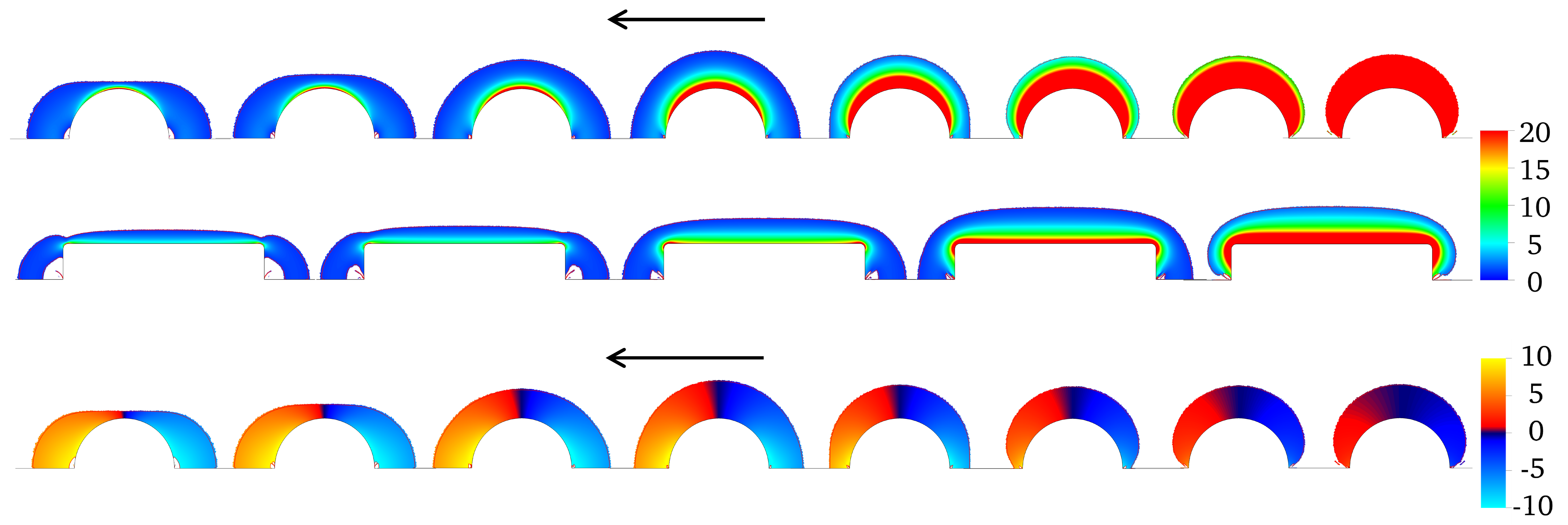}
     \put(0,22.8){\fontsize{7pt}{7pt}\selectfont $\hat{\Omega}=$}
     \put(6,22.8){\fontsize{7pt}{7pt}\selectfont $10^{-1}$}
     \put(19,22.8){\fontsize{7pt}{7pt}\selectfont $10^{-\frac{2}{3}}$}
     \put(31,22.8){\fontsize{7pt}{7pt}\selectfont $10^{-\frac{1}{3}}$}
     \put(44,22.8){\fontsize{7pt}{7pt}\selectfont $10^{0}$}
     \put(55,22.8){\fontsize{7pt}{7pt}\selectfont $10^{\frac{1}{3}}$}
     \put(66,22.8){\fontsize{7pt}{7pt}\selectfont $10^{\frac{2}{3}}$}
     \put(77,22.8){\fontsize{7pt}{7pt}\selectfont $10^{1}$}
     \put(86,22.8){\fontsize{7pt}{7pt}\selectfont $10^{\frac{4}{3}}$}

     \put(0,13.8){\fontsize{7pt}{7pt}\selectfont $\hat{\Omega}=$}
     \put(9,13.8){\fontsize{7pt}{7pt}\selectfont $10^{-1}$}
     \put(28,13.8){\fontsize{7pt}{7pt}\selectfont $10^{-\frac{2}{3}}$}
     \put(46,13.8){\fontsize{7pt}{7pt}\selectfont $10^{-\frac{1}{3}}$}
     \put(65,13.8){\fontsize{7pt}{7pt}\selectfont $10^{0}$}
     \put(83,13.8){\fontsize{7pt}{7pt}\selectfont $10^{\frac{1}{3}}$}

     \put(0,1.8){\fontsize{7pt}{7pt}\selectfont $\hat{\Omega}=$}
     \put(6,1.8){\fontsize{7pt}{7pt}\selectfont $10^{-1}$}
     \put(19,1.8){\fontsize{7pt}{7pt}\selectfont $10^{-\frac{2}{3}}$}
     \put(31,1.8){\fontsize{7pt}{7pt}\selectfont $10^{-\frac{1}{3}}$}
     \put(44,1.8){\fontsize{7pt}{7pt}\selectfont $10^{0}$}
     \put(55,1.8){\fontsize{7pt}{7pt}\selectfont $10^{\frac{1}{3}}$}
     \put(66,1.8){\fontsize{7pt}{7pt}\selectfont $10^{\frac{2}{3}}$}
     \put(77,1.8){\fontsize{7pt}{7pt}\selectfont $10^{1}$}
     \put(86,1.8){\fontsize{7pt}{7pt}\selectfont $10^{\frac{4}{3}}$}

     \put(94,27){\fontsize{7pt}{7pt}\selectfont $\dot{\gamma}$}
     \put(94,12){\fontsize{7pt}{7pt}\selectfont $P$}
    
    \end{overpic}
    \put(-380,130){\makebox(0,0)[lt]{(a)}} 
    \put(-380,47){\makebox(0,0)[lt]{(b)}} 
     \caption{ (a) The shear rate $\dot{\gamma}$ around the sphere and cylinder at various $\hat{\Omega}$ numbers and at a fixed $Bi  = 1$. (b) Pressure field around spherical body for different $\hat{\Omega}$. }
    \label{RoComp1}
 \end{figure}

\HMR{
To elucidate the origin of the transition between these two distinct regimes  reported in Figure~(\ref{fig:FdversusRo_COMP}),}{ we plot the shear rate $\dot{\gamma}$ and pressure distributions at a fixed $Bi = 1.0$ and over a broad range of $\hat{\Omega}$ in Figure~(\ref{RoComp1}). In these visualizations, the sedimentation direction is from right to left (indicated by the arrow). At low $\hat{\Omega}$, large yielded region emerges at the front and rear of the object. However, as $\hat{\Omega}$ increases, rotation induces lateral expansion of the yielded zone, consistent with our PIV measurements. This modification of the yielded region shape results in a sudden modification of the local pressure field. 

This can be seen clearly in $\hat{\Omega} \approx 10^{1/3}$ wherein the expanded midplane yield zone becomes sufficiently large to allow flow circulation from the front to the back of the body. Therefore, the pressure difference between the front and back of the object reduces (Figure~\ref{RoComp1}(b)). The modification of the pressure field at the front and rear of the object explains the rapid decline in total drag observed at higher rotation rates. Overall, the pressure field here is strongly linked to the shape of the yielded region, similar to previous observation by \citet{eastham2022squirmer}. Whether the yield region is controlled by the translational motion or the swirling rotation, the pressure contribution to the drag changes nature and that results in a different power law shown in Figure~(\ref{fig:FdversusRo_COMP}).


}

\subsubsection{Effect of Sedimentation Speed}

\begin{figure}[hthp]
\centering
\includegraphics[width=0.49\linewidth]{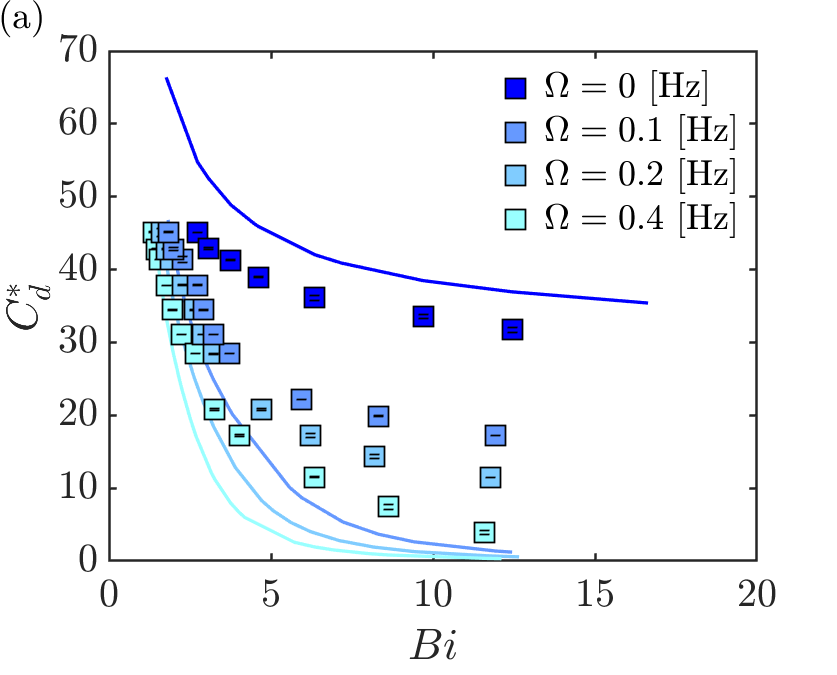}
\includegraphics[width=0.49\linewidth]{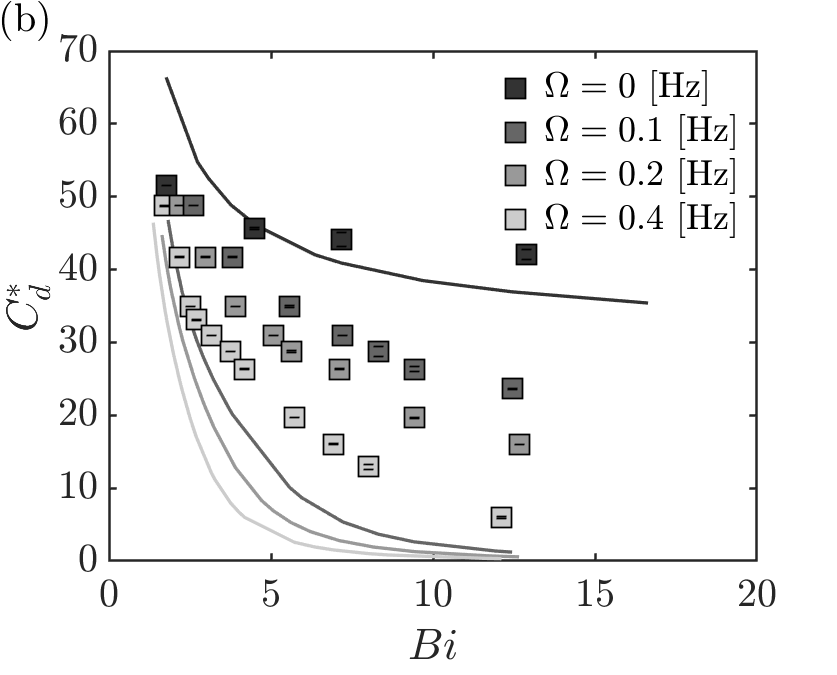}
\includegraphics[width=0.49\linewidth]{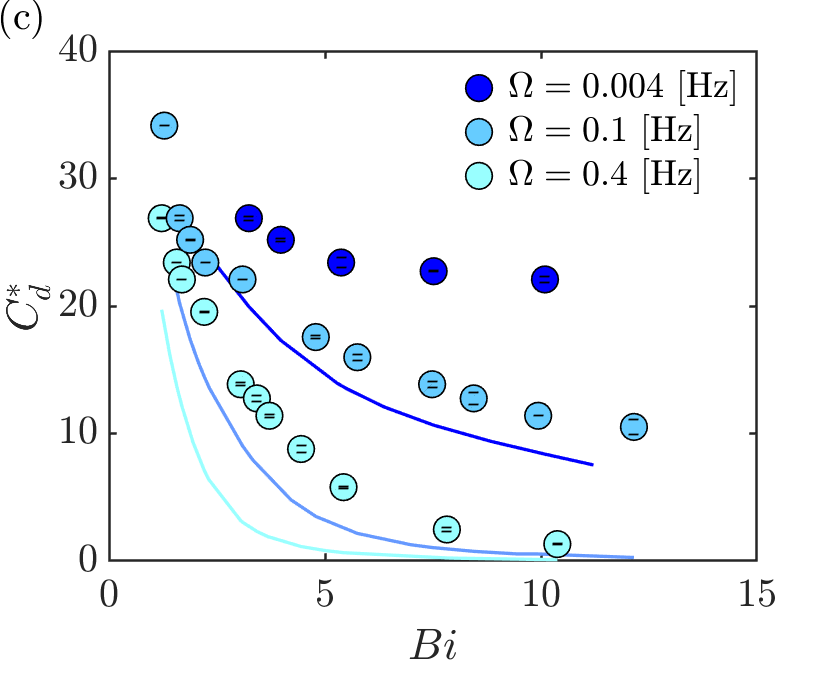}
\includegraphics[width=0.49\linewidth]{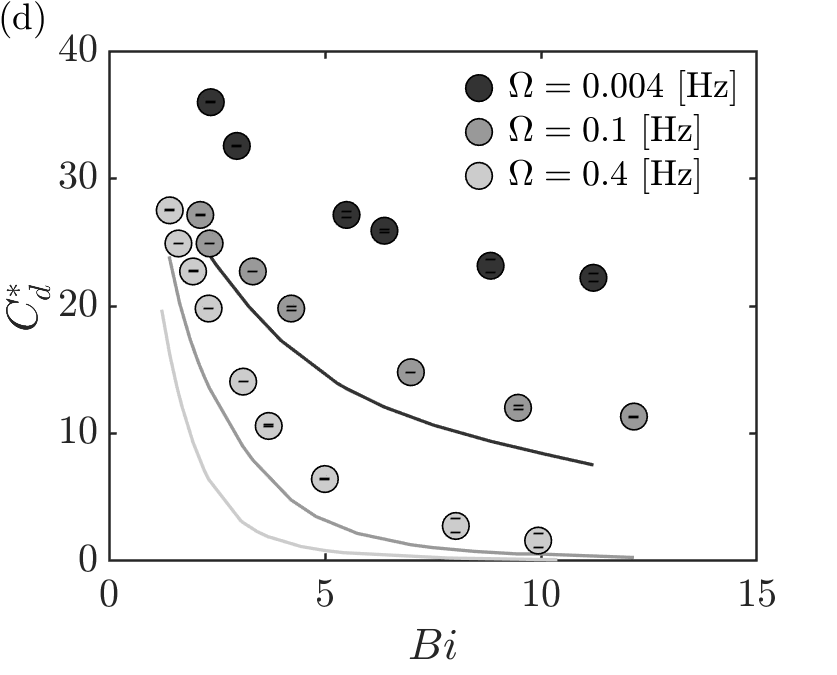}
\caption{Drag coefficient as a function of $Bi$ for different imposed $\Omega$: (a) smooth cylinder, (b) roughened cylinder, (c) smooth sphere, and (d) roughened sphere. Symbols denote experimentally measured values and curves denote the results of numerical simulations. } 
\label{4}
\end{figure}

Figure~(\ref{4}) presents the variation of the measured drag coefficient 
$C^*_d$, as a function of $Bi$ number for different imposed rotational velocities with subfigures (a) and (b) representing smooth and roughened cylinders and (d) and (e) denoting smooth and roughened spheres, respectively. First, at a fixed rotation rate, as $Bi$ number increases, the measured drag coefficient decreases with the drag coefficient being higher for roughened surfaces than the smooth ones. These experimental trends are consistent with existing data reported in the literature when objects do not undergo rotation (or \HM{$\hat{\Omega}=0$)}. Jossic and Magnin \citep{jossic2001drag,jossic2009drag} conducted an experimental study on the settling of smooth and roughened spheres and cylinders in yield stress fluids based on Carbopol 940, and reported that $C^*_d$ follows the same trend as a function of the $Bi$ number. Moreover, they observed that $C^*_d$ is consistently higher for roughened cylinders compared to smooth ones. A similar observation, with a decreasing trend of $C^*_d$ at higher $Bi$, was reported by Merkak et al. ~\citep{merkak2006spheres} for \HM{non rotating} spheres and cylinders by Tokpavi et al.~ \citep{tokpavi2009experimental}.\par

\HM{Secondly,} at a fixed $Bi$ number, our results indicate that rotation produces a lower drag coefficient for all surface types and geometries. \HM{Note, to keep $Bi$ fixed and vary $\Omega$, the mass of the object is systematically and carefully adjusted to maintain a constant sedimentation velocity. The observed reduction in drag with increasing rotation rate is due to the development of a plastic deformation zone that expands around the object as rotation increases. The enlargement of this plastically deformed zone should reduce the overall resistance to motion. }\\

\begin{figure}[hthp]
\centering
\includegraphics[width=0.99\linewidth]{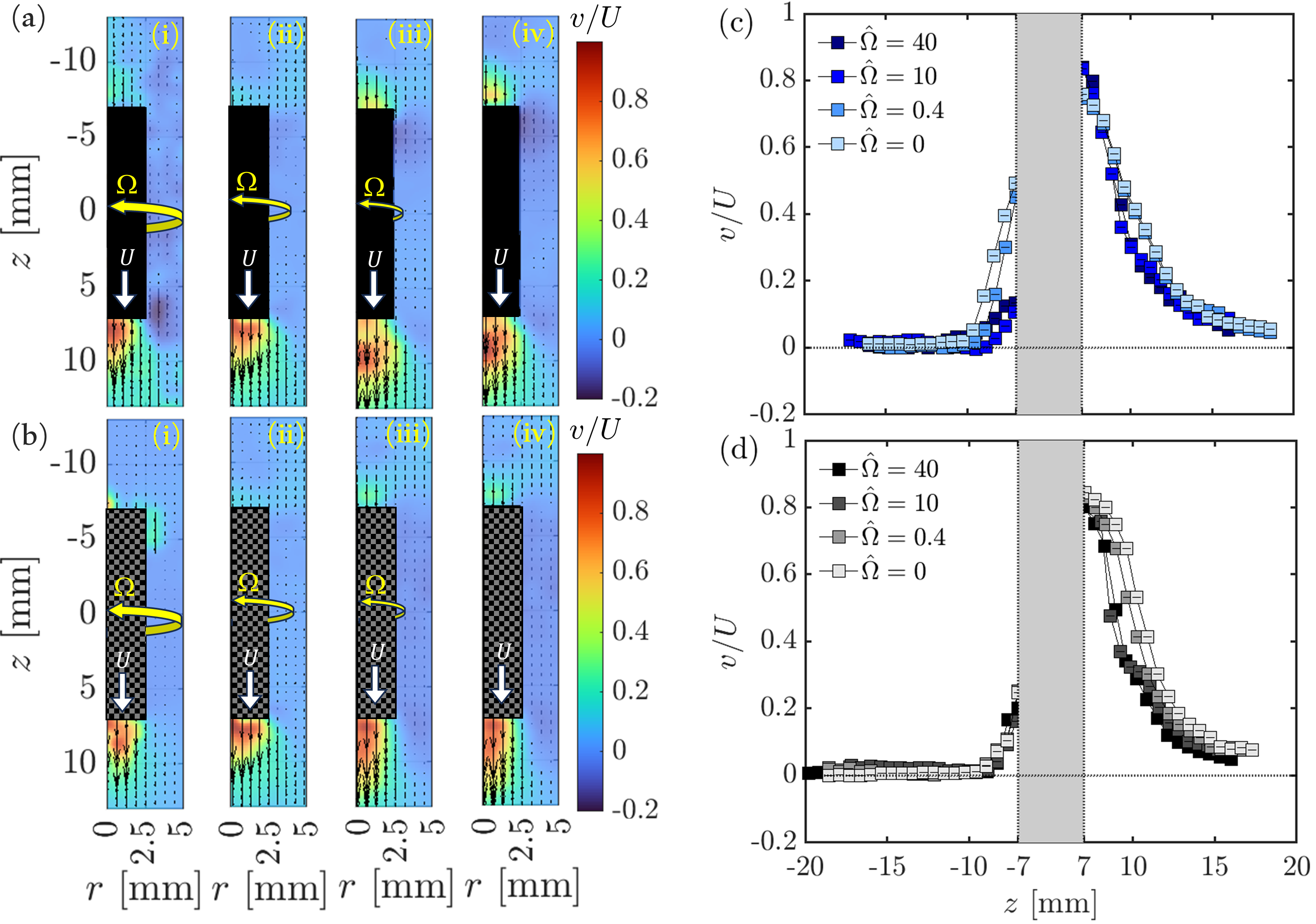}
\caption{2D and 1D averaged normalized velocity profiles in the flow plane for cylinders. (a) Smooth and (b) roughened cylinders at a constant \HM{Bingham number ($Bi\approx 6$) and $\hat\Omega = 40$ (i), 10 (ii), 0.4 (iii), and 0 (iv).} Sub-figures (c) and (d) show the corresponding averaged normalized velocity profiles along the $z$ axis for smooth and roughened cylinders, respectively.}
\label{PIV-FP-cylinder-YS}
\end{figure}

We also performed numerical simulations to evaluate the drag coefficient as a function of $Bi$ for all cases reported in Figure~(\ref{4})\HM{ as solid curves.} In the absence of rotation, the simulations overpredict the drag coefficient for the smooth cylinder, whereas the agreement for the roughened cylinder is noticeably better. For the sphere, measurements without rotation could not be obtained because the required mass exceeded the practical limits imposed by the sphere's size. Once rotation is introduced, the simulations systematically underpredict the experimentally measured drag coefficients as discussed in earlier results as well. To further assess the influence of the Bingham number on the negative wake and the surrounding flow field, we conducted PIV measurements in the flow plane at $Bi \approx6$ and 5 \HM{(constant $U\approx0.025$ [mm/s])} for both cylinders and spheres. Figure~(\ref{PIV-FP-cylinder-YS}) presents the 2D and 1D averaged velocity profiles measured around the cylinders. Interestingly, although the flow field remains asymmetric, the negative wake vanishes even in the absence of rotation. Moreover, increasing the rotation rate does not significantly alter the flow field, regardless of whether the cylinders are smooth or roughened. These findings suggest that the emergence of a negative wake is strongly dependent on the intensity of the flow around the object. In contrast, the flow around spheres at the same $Bi$ still exhibits a negative wake, albeit weaker than that observed at lower Bingham numbers (cf. FIG.~S8 and FIG.~S5 of the SI).\\

\begin{figure}
    \centering
  \vspace{0.5cm}
    \begin{subfigure}[t]{0.49\textwidth}
        \includegraphics[width=\linewidth]{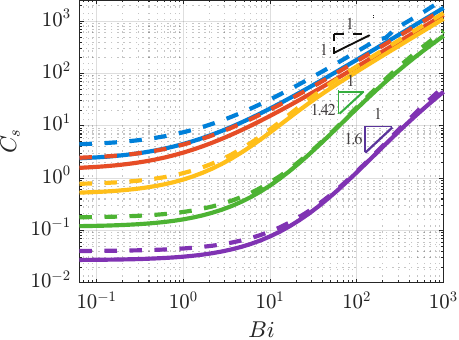}
        \put(-180,155){\makebox(0,0)[lt]{(a)}} 
    \end{subfigure}
    \hfill
    \begin{subfigure}[t]{0.49\textwidth}
        \begin{overpic}[width=1\linewidth,tics=5]{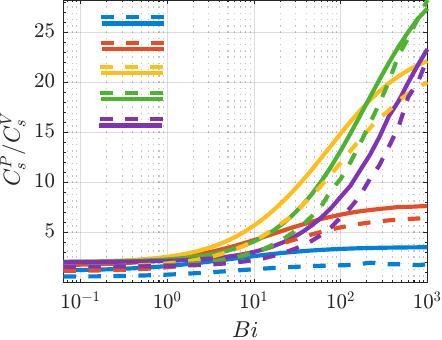}
        \put(40,70){\fontsize{9pt}{11pt}\selectfont $\hat{\Omega} = 0.1$}
       \put(40,65){\fontsize{9pt}{11pt}\selectfont $\hat{\Omega} = 1$}
       \put(40,59){\fontsize{9pt}{11pt}\selectfont $\hat{\Omega} = 10$}
       \put(40,54){\fontsize{9pt}{11pt}\selectfont $\hat{\Omega} = 100$}
       \put(40,48){\fontsize{9pt}{11pt}\selectfont $\hat{\Omega} = 1000$}       
        \end{overpic}
        \put(-180,155){\makebox(0,0)[lt]{{(b)}}}
    \end{subfigure}

    \vspace{0.5em} 

    \begin{subfigure}[t]{0.49\textwidth}
        \includegraphics[width=\linewidth]{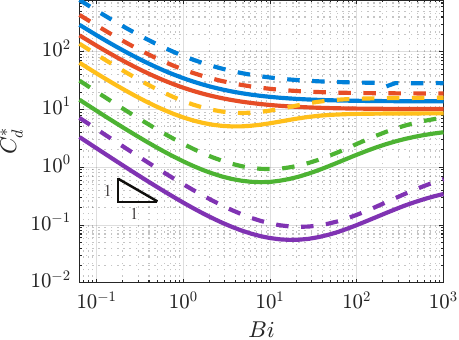}
        \put(-180,155){\makebox(0,0)[lt]{{(c)}}}
    \end{subfigure}
    \hfill
    \begin{subfigure}[t]{0.49\textwidth}
        \includegraphics[width=\linewidth]{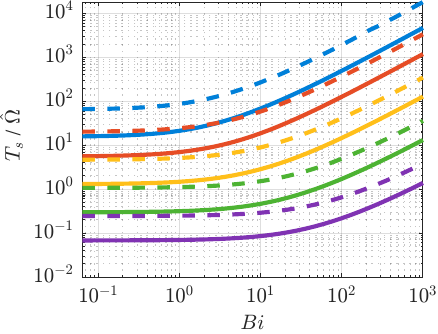}
        \put(-180,155){\makebox(0,0)[lt]{{(d)}}}
    \end{subfigure}

    \caption{Simulation results for (a) normalized drag forces $C_s$, (b) ratio of the pressure drag to shear-induced drag, (c) $C^*_d$ and (d) normalized torque as a function of Rossby number for various $Bi$ numbers. The continuous and dashed curves correspond to simulations on the sphere and the cylinder, respectively.}
    \label{fig:FdversusBi_COMP}
\end{figure}

\begin{figure}[hthp]
\centering
\includegraphics[width=0.9\linewidth]{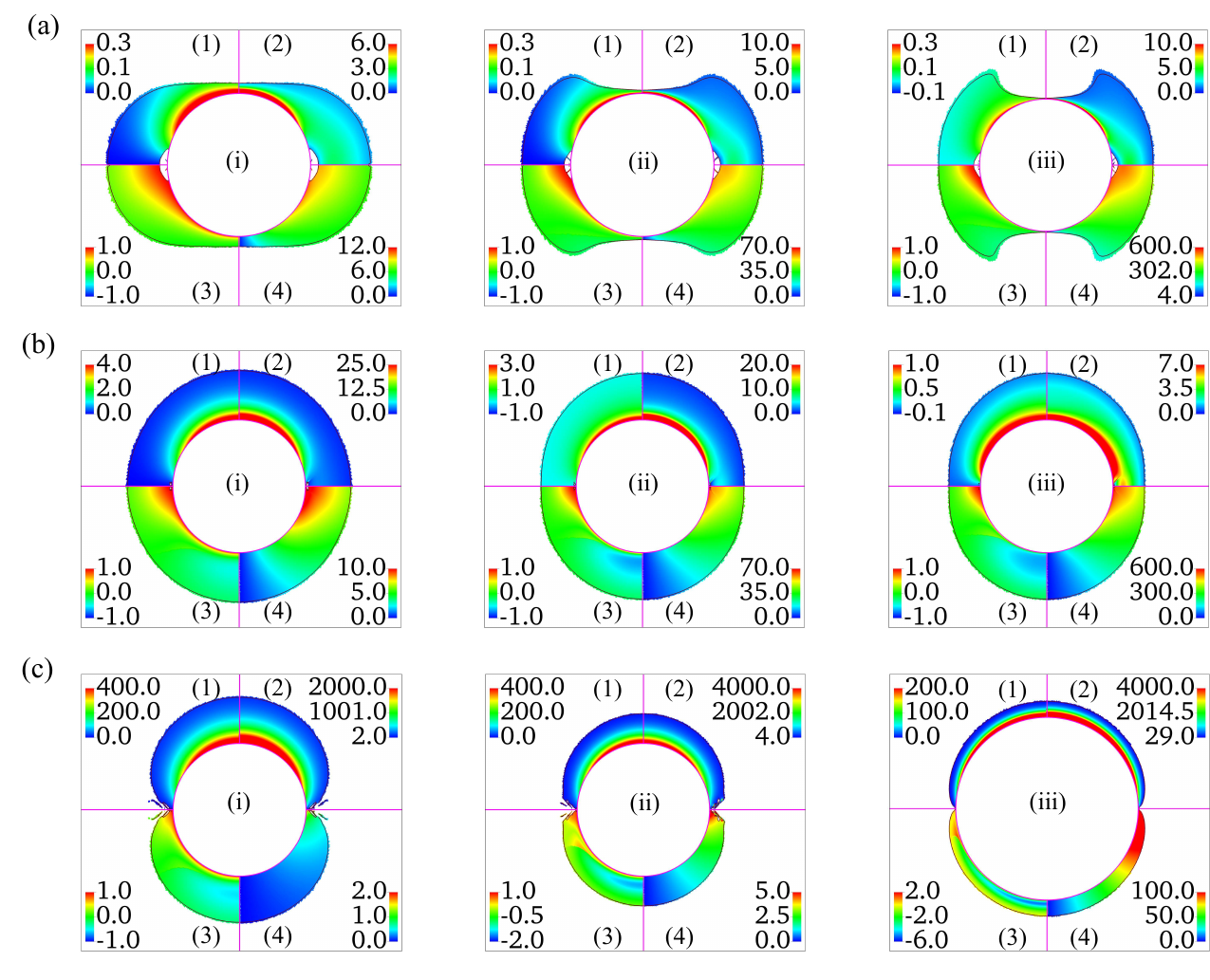}
\caption{Flow field calculated via numerical simulations for $\hat{\Omega} = 0.1$ (a), $1$ (b), and $10$ (c) at three representative Bingham numbers: $Bi = 1$ (\textit{i}), $10$ (\textit{ii}), and $100$ (\textit{iii}). The object sediments from right to left. For each case, the panels represent: (1) normalized azimuthal velocity $u_{\phi}/U$; (2) normalized shear rate $\dot{\gamma}/U$; (3) normalized in-plane velocity magnitude $(\mathbf{|u|}/U) \text{sgn}(u_x)$, where the sign indicates direction relative to sedimentation; and (4) normalized pressure $P/KU^n$.}
\label{RoComp}
\end{figure}

\HM{
To evaluate the sensitivity of the drag coefficient to $Bi$ number, a systematic numerical simulation is performed. Figure~(\ref{fig:FdversusBi_COMP}) shows how $C_s$, $C_s^P/C_s^V$, $C^*_d$ and $T_s$ change as a function of $Bi$ number for a wide range of $\hOmega$ numbers. A similar trend is observed in both the sphere and cylinder cases.

We begin with the behavior of Stokes' drag. For cases with large $Bi$, it is noted that for all $\hOmega < 0.1$, the Stokes drag coefficient ($C_s$) is best scaled by sedimentation velocity and viscosity, showing almost no dependence on the rotational state (see Figure~\ref{fig:FdversusBi_COMP}(a)). Under these conditions, $C_s$  exhibits a slope of approximately $C_s\sim Bi$. These cases correspond to translation-dominated flow fields where the shear rate is strongly localized and the pressure field exhibits a higher value at the frontal and back stagnation regions, resulting in larger plastically deformed (yielded) zones at the front and back of the object and only a narrow yielded layer along its flanks (see Figure~\ref{RoComp}(a)).} \HM{For larger values of $\hOmega$, a $C_s\sim Bi^{\alpha}$, where $\alpha >1 $, and $\alpha$ increases slightly with $\hOmega$. In this regime, the azimuthal velocity induced by rotation generates strong shear near the body surface and expands the surrounding yielded region, allowing the in-plane velocity to circulate more freely around the object (Figure~\ref{RoComp}(b–c)). By contrast, at smaller $Bi$ the variation of $C_s$ is relatively weak, since the rotation largely controls the effective yielded regions surrounding the body.}\par

\HM{The ratio of pressure to frictional viscous drag is shown in Figure~\ref{fig:FdversusBi_COMP}(b). The pressure contribution to the drag coefficient increases as $Bi$ increases. In addition, this ratio exhibits a non-monotonic variation with $\hOmega$ across the entire range of $Bi$. For instance, $\hOmega=10$ produces the largest $C_s^P/C_s^V$ ratio for $Bi\le100$, whereas for larger $Bi$ the maximum shifts to the higher rotational rate $\hOmega=100$. This behaviour results from the evolution of the yielded region generated by the combined translational and swirl motions around the object (Figure~(\ref{RoComp})). As the rotation increases, the azimuthal flow expands the yielded envelope surrounding the body, which modifies the pressure distribution between the front and rear stagnation regions and alters the relative contributions of viscous and pressure drag.}

\HM{Figure~\ref{fig:FdversusBi_COMP}(c) illustrates the variation of the dimensionless drag coefficient $C_d^*$ as a function of the Bingham number $Bi$. For both spherical and cylindrical geometries, two distinct regimes emerge depending on the intensity of the imposed rotation. At low rotation rates, $C_d^*$ decreases as $Bi$ increases and levels off at $Bi>1$. In this regime, the yield surface is governed primarily by sedimentation-driven stresses rather than rotational effects, as illustrated in the flow fields of Figure~\ref{RoComp}(a) and the overall shape of the yield region remains similar. In contrast, for high rotation rates ($\hat{\Omega} \geq 10$, corresponding to rotationally dominated flows), $C_d^*$ exhibits a non-monotonic trend: it initially decreases to a local minimum before rising again at higher $Bi$. This behavior reflects a fundamental transition in the flow topology as can be seen by comparing Figure~\ref{RoComp}(b and c), rapid rotation initially generates an expansive yielded region that envelops the body and reduces the overall drag.

However, as $Bi$ increases, the rotational \HMR{stress} becomes insufficient to overcome the fluid's yield stress. Near the local minimum of $C_d^*$, the yield surface thickness reduces at the front and rear of the body (for example, $Bi \approx 3.16$ for $\hOmega=10$) and eventually becomes pinned, progressively contracting toward the solid boundary. \HMR{Interestingly, in the low-$Bi$ range, a nearly uniform power-law dependence of $C^*_d$ on $Bi$ is observed across all tested $\hat{\Omega}$ curves, where $C^*_d\sim Bi^{-1}$. This is consistent with the observation that as $Bi$ approaches zero, the drag force should be inversely proportional to $Bi$.} At this limit, different rotation rates primarily modulate the spatial extent of the already enlarged yield domain. By altering this effective confinement, the rotation modifies the resulting hydrodynamic resistance while its relation to $Bi$ remains almost unchanged. \HMR{The physical reason behind the observed local minimum in $C_d^*$ is discussed in Appendix~A.}}

\begin{figure}[hthp]
  \centering
    \includegraphics[width=1\linewidth]{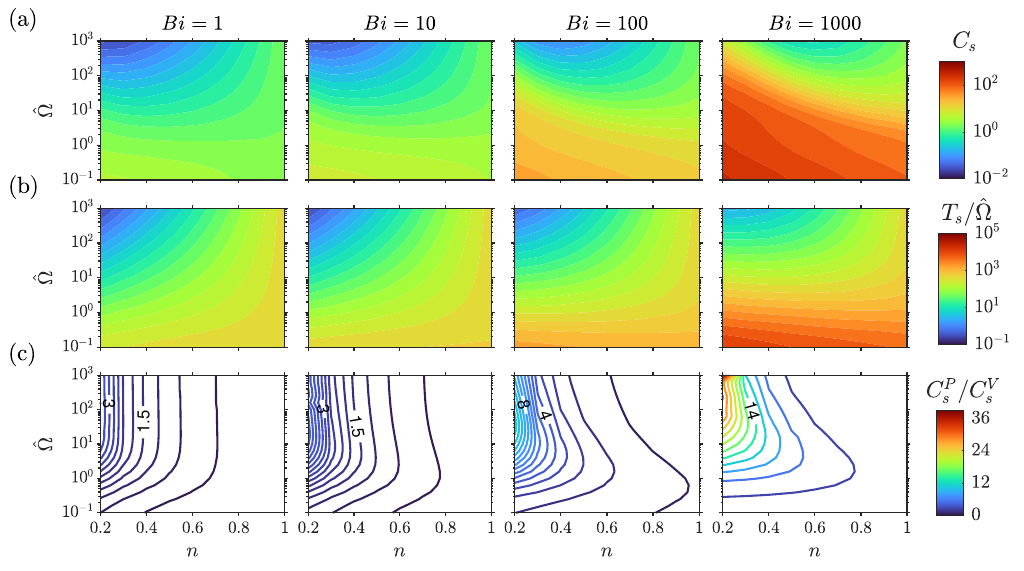}
  \caption{The heat map that shows the prediction of nondimensional drag $C_s$ (a), torque $T_s$ (b) and the ratio of pressure drag to viscous drag $C_s^P/C_s^V$ (c) as a function of $\hat{\Omega}$ number and shear thinning index $n$ for the cylinder with $L/D=2.8$}
  \label{fig:Roversusn_COMP_cylinder}
\end{figure}

\HM{
\subsection{Shear-Thinning Effects}

The drag on a sedimenting and rotating object is expected to depend not only on the yield stress but also on the fluid’s shear-thinning behavior. Although the effect of the shear-thinning power-law index on the plastic drag coefficient was not investigated experimentally, we examined it systematically through numerical simulations. For $n=1$, the fluid behaves as a Bingham plastic, whereas for $n<1$, the apparent viscosity decreases with increasing shear rate according to a power-law relationship. Figure~(\ref{fig:Roversusn_COMP_cylinder}) presents contour maps of the dimensionless Stokes drag $C_s$, torque $T_s$, and the pressure-to-viscous drag ratio $C_s^P/C_s^V$ for the cylinder as functions of the dimensionless rotation rate $\hOmega$ and the shear-thinning index $n$, for four representative values of $Bi$. The corresponding plots for the sphere are shown in the FIG.~S9 of the SI.}\par  

\HM{In principle, simulations suggest two distinct regimes based on the topology of the yielded zone: a) rotationally-induced yielding (high $\hOmega$), and b) translationally-induced yielding (small $\hOmega$). In the rotation-dominated regime, the rapid rotational motion generates intense local shear rates in the immediate vicinity of the object's surface. Because the shear rate is elevated, the fluid's apparent viscosity in this rotationally induced yield region becomes exceptionally sensitive to $n$. Enhancing the shear-thinning effect (decreasing $n$) sharply reduces the local effective viscosity, effectively creating a highly mobilized, lubricated layer that envelops the body. Consequently, \HMR{as shown in Figure~\ref{fig:Roversusn_COMP_cylinder}(a,b)} both the drag $C_s$ and torque $T_s$ exhibit their highest sensitivity to $n$ in this limit. Interestingly, for very strong rotation rates (e.g., $\hat{\Omega}>100$) the complex coupling between the yielded envelope size and the lubricated boundary layer leads to non-monotonic variations in the Stokes' drag coefficients with respect to $n$.}\par

\HM{Conversely, when translation dominates ($\hat\Omega \leq1$), the yield surface and the local shear-rate distributions are primarily dictated by the sedimentation of the object. Here, shear is concentrated primarily at the equator, driven by the front-to-back displacement of the fluid. Because the intense shear is less uniformly distributed around the body than in the rotational case, the overall hydrodynamic resistance shows a weaker, more monotonic dependence on the shear-thinning index. Still, as $n$ increases toward Newtonian behaviour, the drag increases steadily but with a lower rate than rotationally dominated flows, reflecting the loss of the shear-thinning effect. Additionally, the contour maps of $ C_s^P/C_s^V $ (shown in Figure~\ref{fig:Roversusn_COMP_cylinder}(c)) illustrate that shear-thinning significantly alters the balance of pressure and viscous drag forces acting on the object. The localized decrease in viscosity caused by shear-thinning (i.e., lower $n$ values) leads to an increased relative contribution of the pressure form drag, especially in areas where rotation is significant.}\par

\subsection{Yield Limit}
An important aspect of our experiments is that, below a certain critical mass (or buoyant force), the object ceases to sediment and effectively becomes trapped within the yield-stress medium. In the following, we explore how introducing rotational motion can alter this threshold and trigger sedimentation. Central to this analysis is establishing an experimental criterion for identifying the onset of sedimentation. In the experiments reported here, the yield limit is determined by measuring the settling velocity as a function of added weight to the object. The object’s weight is incrementally adjusted until no detectable sedimentation occurs within a 20 min observation period (see FIG.~S10 in the SI for object trajectories over time at different added weights). Note that experimental duration cannot be extended beyond 20 min due to overheating of the Helmholtz coil. The transition occurs from no motion (within 20 min) to slight motion with $U \approx 0.1 $ [mm/min]. Moreover, the mass difference between these two conditions is minimal ($\leq$0.01 g), which approaches the resolution limit of our analytical balance (1 mg), making precise adjustments below 0.01 g challenging and less reliable. \HM{Therefore, the onset of sedimentation is defined as the maximum object mass for which no measurable translation is observed over the 20-minute duration of the experiment \HMR{(i.e., $U = 0$)}.} \par 

\begin{figure}[hthp]
\centering
\includegraphics[width=0.48\linewidth]{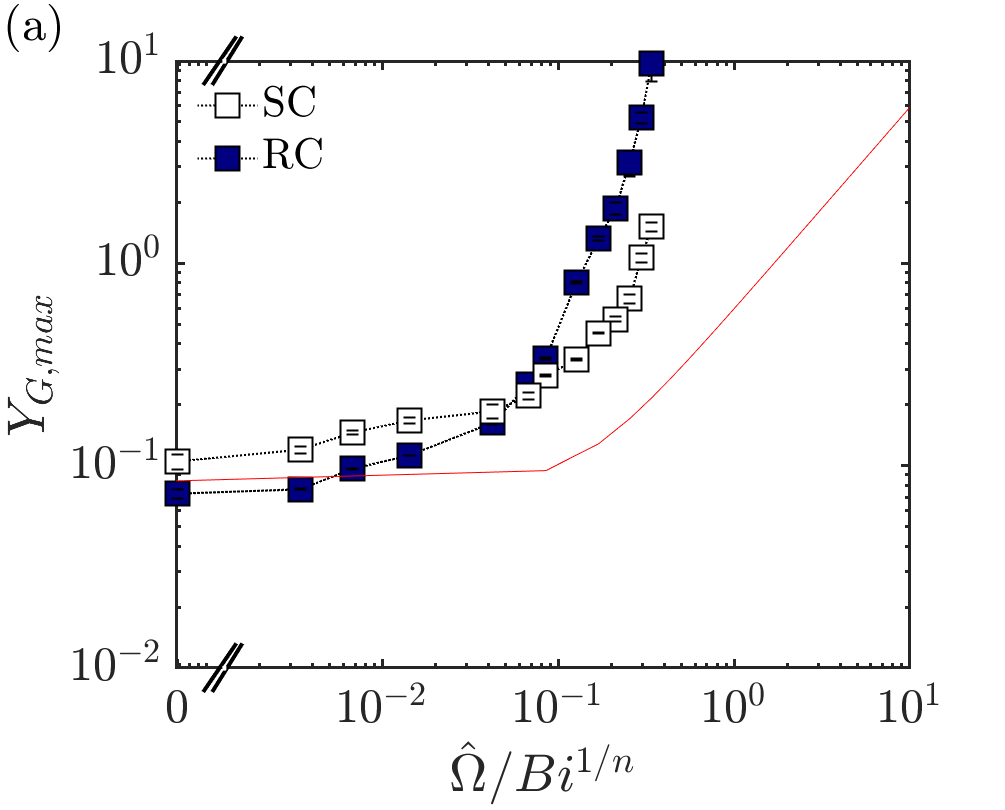}
\includegraphics[width=0.48\linewidth]{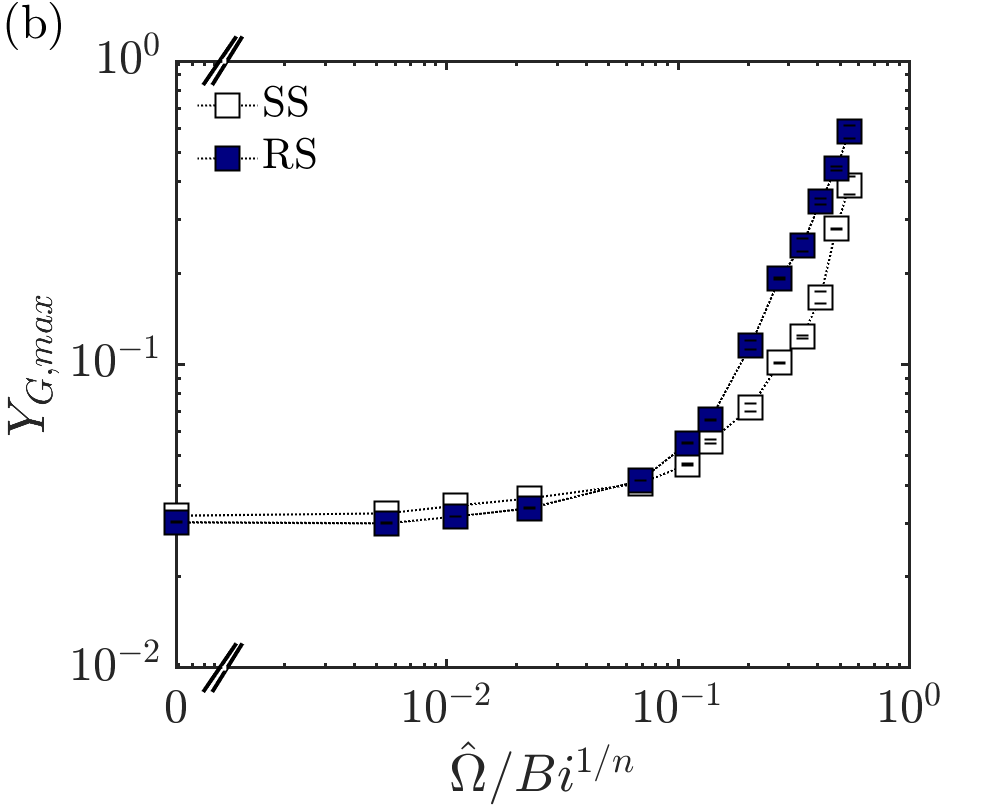}

\caption{Dimensionless \HM{yield limit as a function of $\hat{\Omega}/Bi^{1/n}$ } for smooth and roughened cylinders (a) and spheres (b) in the yield stress fluid. Here, SC and RC refer to smooth and roughened cylinders. In addition, SS and RS are smooth and roughened spheres. \HMR{
The continuous curve in part~(a) shows the high-$Bi$ infinite-cylinder scaling
of Hewitt and Balmforth~\citep{hewitt2018viscoplastic}. This curve is included
as a qualitative asymptotic reference, not as a finite-cylinder fit.
}}
 
\label{Yield limit}
\end{figure}

Figure~(\ref{Yield limit}) presents the measured yield limit as a function of imposed $\Omega$ \HM{(in this case $\hat{\Omega}/Bi^{1/n}$)} for both spherical and cylindrical objects with smooth and roughened surfaces. \HMR{In defining the yield limit, we use conditions under which $U = 0$. To eliminate the dependence of the dimensionless groups $Bi$ and $\hat{\Omega}$ on the translational velocity $U$, we therefore choose to present the x-axis using the combined dimensionless group $\hat{\Omega} / Bi^{1/n} $. This rescaling removes the explicit dependence on $U$ at the yield limit. } Several key observations emerge from these experiments. First, the measured yield limit decreases with decreasing rotation for both the cylinder (Figure~\ref{Yield limit}(a)) and the sphere (Figure~\ref{Yield limit}(b)), eventually leveling off for \HM{$ \hat{\Omega}/Bi^{1/n} < 10^{-1}$}. Second, across nearly all tested rotational velocities, the yield limit is consistently higher for the cylinder than for the sphere. Third, surface roughness plays a significant role: at low \HM{$\hat\Omega/Bi^{1/n}$} numbers, roughened surfaces exhibit a lower yield limit, whereas at higher \HM{$\hat\Omega/Bi^{1/n}$} values ( \HM{$\hat\Omega/Bi^{1/n}>0.1$}), smooth surfaces show lower yield limits.  \\

Although there is no \HMR{experimentally obtained} data in the published literature on the effect of rotation on the yield limit, there are several reports on yield limit values in the absence of rotation. For the non-rotating limit \HM{($\hat{\Omega}/Bi^{1/n} =0$)}, our measured yield limit \HM{values ($Y_{G,max} = 0.07$ for RC, $Y_{G,max} = 0.11$ for SC, $Y_{G,max} = 0.03$ for RS, and $Y_{G,max} = 0.035$ for SS)} \HMR{are slightly different from the} published results. For example, Jossic and Magnin~\citep{jossic2001drag} reported \HM{$Y_{G,max} \approx 0.062$ for a roughened sphere and $Y_{G,max} \approx 0.088$ for a smooth sphere.} Consistent with our experiments, their results also indicate that the yield limit is higher for smooth surfaces. Other experimental studies have reported yield limits of approximately 0.048 and 0.061 for spheres in yield stress fluids~\citep{atapattu1995creeping,tabuteau2007drag}. Likewise, the asymptotic analysis of Beris et al.~\citep{beris1985creeping}, using a Bingham fluid model, estimated \HM{$Y_{G,max} \approx 0.048$ for a sphere.} Similar consistency is observed for falling cylinders. For cylinders oriented along their longest axis, the yield limit depends on both the aspect ratio and surface roughness. Jossic and Magnin~\citep{jossic2001drag} measured \HM{$Y_{G,max} \approx 0.15$ for a smooth cylinder and $Y_{G,max} \approx 0.075$ for a roughened cylinder with an aspect ratio of 5.} Numerical work by Iglesias~\citep{iglesias2020computing} likewise reported comparable yield limit values for cylinders with varying aspect ratios. \HMR{The observed discrepancy between the yield-limit values obtained in our experiments and those reported in the literature may be due to variations in cylinder aspect ratio, surface roughness, and/or the rheological characteristics of the surrounding fluid.}\\

\begin{figure}[h]
\centering
\includegraphics[width=0.7\linewidth]{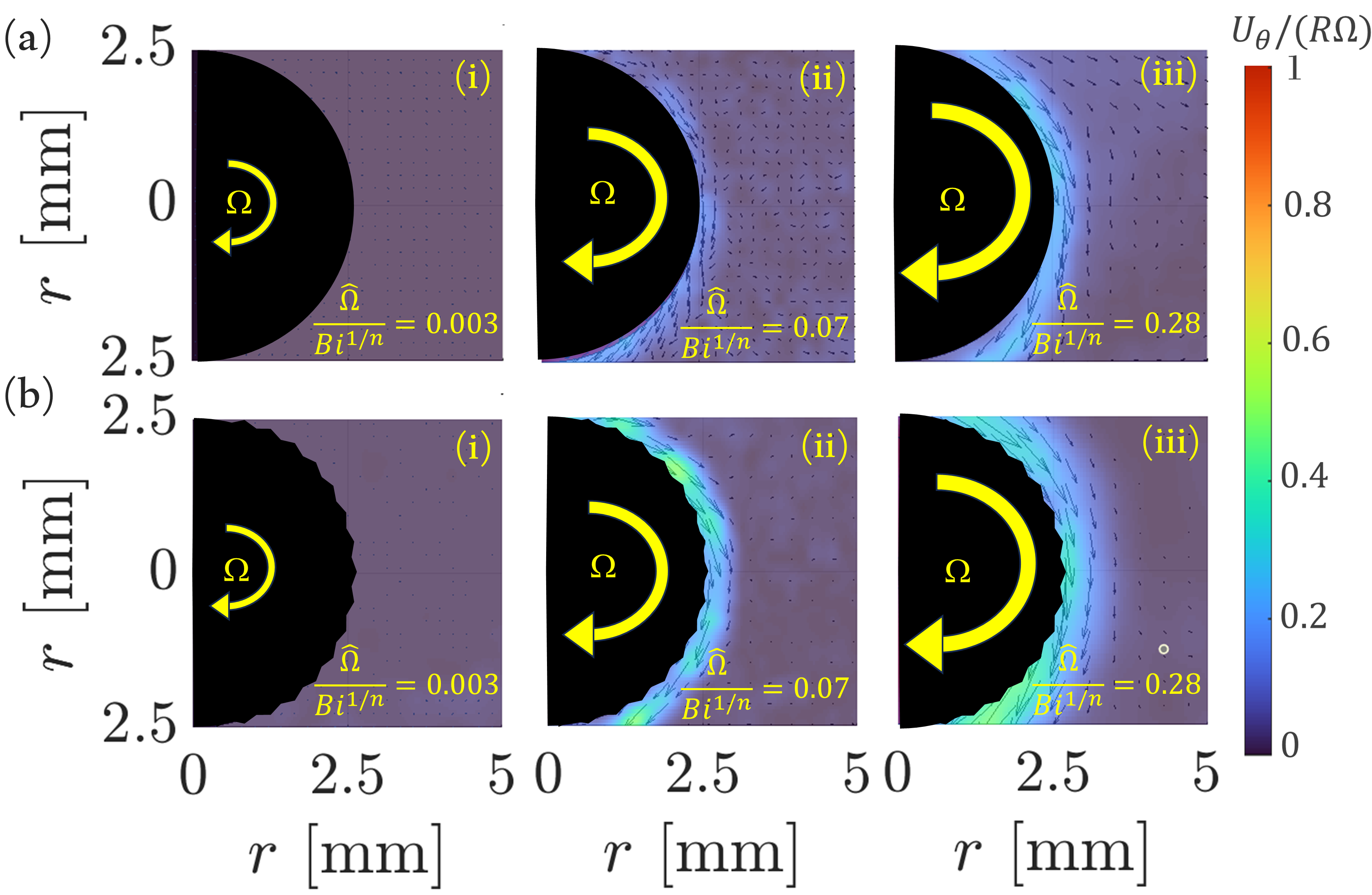}
\includegraphics[width=0.5\linewidth]{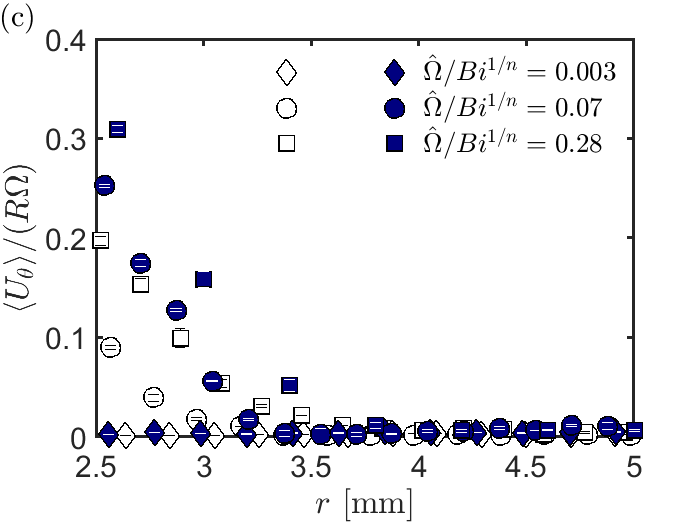}

\caption{2D averaged velocity profile around the cylinders. Smooth (a) and roughened (b) with various \HM{$\hat{\Omega}/Bi^{1/n}$ values from left to right (0.003, 0.07 and 0.28).} (c) Radially averaged normalized velocity magnitude for cases shown in (a), and (b). In (c), the filled and open symbols denote the results for roughened and smooth surfaces, respectively. } 
\label{6}
\end{figure}

\HMR{The above experiments indicate that rotation has little effect on the yield limit at low imposed rotations. At higher rotation rates, however, $Y_{G,\max}$ increases markedly, indicating that rotation facilitates the onset of sedimentation in yield-stress fluids. 
This trend can be interpreted through the combined shear and pressure contributions to the drag. 
Rotation generates strong azimuthal shear near the particle surface and produces a yielded envelope around the body. 
In the high-swirl limit, the surface traction becomes increasingly aligned with the azimuthal direction, so that only its meridional projection contributes to the axial resistance. 
The shear contribution to the plastic drag therefore decreases as
\begin{equation}
C^{*}_{d,\tau}
\sim
C_{\tau}\left(1+\hat{\Omega}^{2}\right)^{-1/2}.
\end{equation}
If the drag were purely shear-dominated, this would imply
\begin{equation}
Y_G
\sim
\left(1+\hat{\Omega}^{2}\right)^{1/2},
\end{equation}
consistent with the high-$Bi$ infinite-cylinder result of~\citep{hewitt2018viscoplastic}. 
For finite objects, however, Appendix~A shows that confinement of the yielded layer can also produce a pressure contribution,
\begin{equation}
C^{*}_{d,p}
\sim
C_p Bi^{2/(n+1)}
\hat{\Omega}^{-(3n+1)/(n+1)}.
\end{equation}
Thus, the relevant resistance is
\begin{equation}
C_d^*
=
C^{*}_{d,\tau}
+
C^{*}_{d,p},
\qquad
Y_G \sim \frac{1}{C_d^*}.
\end{equation}
The observed increase in $Y_{G,\max}$ therefore reflects a reduction in the axial shear resistance by rotation, while deviations from the shear-only scaling can arise from pressure drag associated with finite geometry and yielded-layer confinement.

To examine this interpretation experimentally, we performed detailed PIV measurements around the cylinder and sphere near the onset of sedimentation. 
Figure~\ref{6} shows the flow field around smooth and roughened cylinders at different values of 
$\hat{\Omega}/Bi^{1/n}$ at the onset of sedimentation. 
At small values of $\hat{\Omega}/Bi^{1/n}$, the surrounding material barely deforms, and no clear plastic deformation zone is captured within the spatial resolution of the PIV measurements. 
At $\hat{\Omega}/Bi^{1/n}=0.07$, a thin plastic deformation zone becomes visible near the rotating object. 
As $\hat{\Omega}/Bi^{1/n}$ increases further, this yielded zone grows and moves farther away from the cylinder surface. 
Similar behavior is observed for the sphere, as shown in Fig.~S11 of the SI. 
These observations support the idea that the increase in $Y_{G,\max}$ at high rotation is associated with the formation and expansion of a rotation-induced yielded region around the object.}
\\

Another feature of experimental results reported in Figure~(\ref{Yield limit}) is that the roughened objects have higher yield limits compared to the smooth surfaces at high rotation rates \HM{ (or high $\hat\Omega/Bi^{1/n}$} values). This suggests that onset of sedimentation requires less mass for roughened surfaces. These results are presumably correlated with the plastic deformation zone and/or wall-slip behavior around the rotating objects. Our PIV results of Figure~\ref{6}(b) demonstrate that at high \HM{$\hat\Omega/ Bi^{1/n}$} numbers, the wall slip is less pronounced around the roughened cylinder. The larger plastically deformed zone around roughened cylinder, which may be caused by less wall-slip, may create a corridor of smaller resistance and lead to \HM{a higher yield limit (or equivalently less mass needed to achieve onset of motion).} \\

\begin{figure}[hthp]
 \centering
 \includegraphics[width=0.5\linewidth]{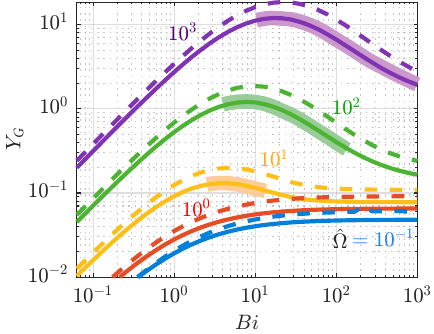}
\caption{The gravitational yield number as a function of Bingham number for various imposed rotations ($\hat{\Omega}$). Continuous curves correspond to the sphere, and dashed curves correspond to the cylindrical cases. The highlighted segments mark the high-swirl thin-layer regime. At fixed $\hat{\Omega}$, increasing $Bi$ thins the rotationally yielded layer, and strengthens the pressure contribution to drag. Because $Y_G$ is
inversely proportional to drag, the numerical curves decrease after the peak in the highlighted ranges. Larger $\hat{\Omega}$ shifts the valid window to larger $Bi$ showing  clear consistency with the scaling.} 
\label{simulationYG}
\end{figure}

\HMR{ We note that the numerical simulations should be interpreted with caution in the context of yield limits. The experiments define the yield limit operationally as complete
arrest within experimental resolution, $U=0$. By contrast, the simulations
impose a finite sedimentation velocity and therefore approach the yield limit
only asymptotically as $Bi$ becomes large. This distinction already exists for
non-rotating particles, where the computed resistance approaches the critical
condition discussed by \cite{beris1985creeping}, but does not directly represent a
static arrest calculation. For rotating bodies, the high-$Bi$ infinite-cylinder
solution of~\citep{hewitt2018viscoplastic} predicts a limiting force
${F_d}/{Bi}
\sim
{2\pi}{\sqrt{1+\hat{\Omega}^{2}}}$,
implying
$
Y_G
\sim
\sqrt{1+\hat{\Omega}^{2}}$ in the limit of very high $Bi$ numbers.
The comparison shown in Fig.~\ref{Yield limit}(a) indicates that the experimental cylinder
data follow the same increasing trend with rotation, although quantitative
agreement is not expected because the present cylinders are finite, the
experiments may involve wall slip, and pressure drag becomes important at high
rotation and high $Bi$. Appendix~A provides an analogous local scaling argument
for a rotating sphere, showing that the shear contribution gives the same
$\sqrt{1+\hat{\Omega}^{2}}$ increase in $Y_G$, while a confined
yielded layer introduces an additional pressure contribution. The latter
explains why the apparent $Y_G$ inferred from velocity-controlled simulations
can vary non-monotonically with $Bi$. Therefore, the simulations and scaling
arguments are used here to interpret trends in the onset of motion, not to
claim an exact numerical prediction of the arrested yield limit. 

It is worth noting that the comparison between the simulation-derived $Y_G$ and
the scaling in Appendix~A should be restricted to the portions of the curves
that satisfy the asymptotic validity window. For the present Herschel--Bulkley
index, $n=0.36$, this window is
\[
\hat{\Omega}^{0.36}\ll Bi \ll \hat{\Omega}^{1.04}.
\]
The highlighted intervals in Fig.~\ref{simulationYG} identify where this
condition is approximately satisfied for each imposed rotation. Within these
intervals, the high-rotation numerical curves follow the trend predicted by the
local scaling: after the rotation-assisted maximum, $Y_G$ decreases with
increasing $Bi$ as the yielded layer becomes increasingly confined and the
pressure contribution becomes more important. In the pressure-dominated limit,
the predicted dependence for $n=0.36$ is
\[
Y_G\propto Bi^{-1.47}\hat{\Omega}^{1.53},
\]
which is consistent with the downward branches observed for the
$\hat{\Omega}=10^2$ and $\hat{\Omega}=10^3$ curves. The lower-rotation cases and
the portions of the curves outside the highlighted windows are not expected to
follow this asymptotic trend quantitatively, because the assumptions of
rotation-dominated yielding and a thin near-surface yielded layer are no longer
fully satisfied. Moreover, because the simulations impose a finite settling
velocity, they provide an apparent $Y_G$ based on the computed drag rather than
the unique arrested yield limit $Y_{G,\max}$ measured experimentally.}

\section{Conclusion and Outlook}
We investigated the sedimentation of rotating objects in a yield-stress fluid using a combination of experiments and numerical simulations. The results of this study can be summarized as follows:\\
The drag coefficient was found to be highly sensitive to rotation rate. Our results indicate that increasing \HM{rotation rate or $\hat{\Omega}$} lead to lower drag for both cylinders and spheres, with roughened surfaces consistently exhibiting greater drag than smooth ones. Numerical simulations reproduced the general trends in drag coefficient with respect to \HM{$\hat{\Omega}$}, but systematically underpredicted the measured drag, likely due to wall slip and nonlinear flow features such as the formation of negative wakes behind the falling object. Furtheremore, increasing the \HM{Bingham number} reduced the drag coefficient, reflecting the growing dominance of viscous forces over yield stress. The scaling laws are identified based on simulations over a wide range of \HM{$Bi$ and $\hat{\Omega}$.} \\
We further examined the yielding criterion and its dependence on rotation. \HM{The results reveal that the yield limit decreases significantly with decreasing rotation rate $\hat{\Omega}$}, but levels off for \HM{$\hat\Omega/Bi^{1/n}< 0.1$. Surface roughness also plays a central role: at low $\hat{\Omega}$, smooth surfaces exhibit a higher yield limit, whereas at high $\hat{\Omega}$ roughened surfaces show a higher yield limit.} \HMR{The increase in the measured yield limit with rotation is consistent with
high-$Bi$ theory for a rotating/translating cylinder and with the lubrication
scaling in Appendix~A: rotation reduces the axial projection of the shear
resistance, while finite-body pressure effects and wall slip can modify the
quantitative threshold.} Finally, from a computational perspective, future improvements to the predictive model could include adopting an elastoviscoplastic framework and incorporating surface slip conditions.
\section{Acknowledgements}
We gratefully acknowledge partial support for this work through National Science Foundation award CBET 2512810.

\HMR{}{
\appendix
\section{Lubrication scaling for high-swirl drag and gravitational yield number}
\label{sec:lubrication_YG_high_swirl}

The simulations show that, at large $\hOmega$, the relative contributions of
pressure and shear stresses to the drag are strongly affected by the
structure of the yielded region surrounding the particle. We next develop a
compact lubrication estimate to rationalize this behavior and to connect the
high-swirl drag scaling to the gravitational yield number. The analysis is
intended as a local scaling argument rather than a closed solution of the full
free-boundary viscoplastic problem.

Consider a sphere of radius $R$ translating with velocity $U\bm{e}_z$ and
rotating about the same axis with angular velocity $\Omega\bm{e}_z$. In
spherical coordinates $(r,\theta,\phi)$, where $\theta$ is measured from
the settling direction, the velocity imposed on the particle surface is
\begin{equation}
u_r^s=U\cos\theta,
\qquad
u_\theta^s=-U\sin\theta,
\qquad
u_\phi^s=\Omega R\sin\theta .
\label{eq:surf_vel_lub_final}
\end{equation}
The tangential surface velocity is therefore
\begin{equation}
\bm{u}_t^s
=
-U\sin\theta\,\bm{e}_\theta
+
\Omega R\sin\theta\,\bm{e}_\phi,
\qquad
|\bm{u}_t^s|
=
U\sin\theta(1+\hOmega^2)^{1/2},
\label{eq:tangential_vel_lub_final}
\end{equation}
where $\hOmega=R\Omega/U$. In the high-swirl regime, $\hOmega\gg1$, the local
shear is dominated by the azimuthal motion, while the meridional component
associated with sedimentation gives a smaller projection onto the settling
direction.

At large $Bi$, the rotationally yielded region may be localized near the
particle surface. We denote its outer boundary by

\begin{equation}
r_y(\theta)=R+h(\theta),
\qquad
h(\theta)\ll R,
\label{eq:thin_layer_lub_final}
\end{equation}
and introduce the local normal coordinate $y=r-R$, with
$0<y<h(\theta)$. In this thin layer, the dominant rotational shear rate is
estimated as

\begin{equation}
\dot{\gamma}_\Omega
\sim
\frac{\Omega R\sin\theta}{h}.
\label{eq:gammaOmega_lub_final}
\end{equation}
For a Herschel--Bulkley material, the corresponding viscous stress scale is
\begin{equation}
\tau_\Omega^v
\sim
k
\left(
\frac{\Omega R\sin\theta}{h}
\right)^n .
\label{eq:rot_stress_lub_final}
\end{equation}
The yielded layer terminates where the stress magnitude approaches the yield
stress. For a rotating sphere, the azimuthal shear stress decreases away from
the surface because of curvature. Thus, the excess wall stress needed to sustain
a yielded layer of thickness $h$ scales as
\begin{equation}
\tau_\Omega^v
\sim
C_\kappa\sigma_y\frac{h}{R},
\label{eq:curvature_balance_lub_final}
\end{equation}
where $C_\kappa=O(1)$ is a geometric curvature constant (e.g. for purely
azimuthal Stokes flow around a sphere, $C_\kappa=3$). Combining
Eqs.~(\ref{eq:rot_stress_lub_final}) and
(\ref{eq:curvature_balance_lub_final}) gives
\begin{equation}
\frac{h(\theta)}{R}
\sim
\left[
\frac{\sin^n\theta}{C_\kappa Bi_\Omega}
\right]^{1/(n+1)},
\label{eq:h_BiOmega_lub_final}
\end{equation}
where the rotation-based Bingham number is $Bi_\Omega=Bi/\hOmega^n$. This thin-layer approximation
requires $Bi_\Omega\gg1$, or equivalently $\hOmega^n\ll Bi$.

\subsection{Shear contribution}
Let $\bm{t}_\tau=\boldsymbol{\tau}\cdot\bm{n}$ denote the deviatoric or shear
traction acting on the particle surface. In the large-$Bi$ thin-layer limit, the
magnitude of the plastic part of this traction is approximately $\sigma_y$, with
a smaller Herschel--Bulkley correction. Using
Eq.~(\ref{eq:curvature_balance_lub_final}),
\begin{equation}
|\bm{t}_\tau|
\sim
\sigma_y
+
k
\left(
\frac{\Omega R\sin\theta}{h}
\right)^n
=
\sigma_y
\left[
1+O\left(\frac{h}{R}\right)
\right].
\label{eq:traction_mag_lub_final}
\end{equation}
The tangential traction opposes the local surface motion, so that
\begin{equation}
\bm{t}_\tau
\sim
-
|\bm{t}_\tau|
\frac{\bm{u}_t^s}{|\bm{u}_t^s|}.
\label{eq:traction_vector_lub_final}
\end{equation}
Only the meridional component of $\bm{t}_\tau$ contributes to the axial drag,
because $\bm{e}_\phi\cdot\bm{e}_z=0$ and
$\bm{e}_\theta\cdot\bm{e}_z=-\sin\theta$. Hence,
\begin{equation}
-\bm{t}_\tau\cdot\bm{e}_z
\sim
|\bm{t}_\tau|
\frac{\sin\theta}{(1+\hOmega^2)^{1/2}}.
\label{eq:traction_projection_lub_final}
\end{equation}
The shear contribution to the drag is therefore
\begin{equation}
F_{d,\tau}
=
-\int_S \bm{t}_\tau\cdot\bm{e}_z\,dS
\sim
2\pi R^2\sigma_y(1+\hOmega^2)^{-1/2}
\int_0^\pi \sin^2\theta\,d\theta .
\label{eq:shear_drag_int_lub_final}
\end{equation}
Using $\int_0^\pi \sin^2\theta\,d\theta=\pi/2$, we obtain
\begin{equation}
F_{d,\tau}
\sim
\pi^2R^2\sigma_y(1+\hOmega^2)^{-1/2}.
\label{eq:shear_drag_lub_final}
\end{equation}
For a sphere, $A=\pi R^2$, so the shear contribution to the plastic drag
coefficient is
\begin{equation}
C_{d,\tau}^*
=
\frac{F_{d,\tau}}{\pi R^2\sigma_y}
\sim
\pi(1+\hOmega^2)^{-1/2}
\sim
\pi\hOmega^{-1},
\qquad
\hOmega\gg1 .
\label{eq:Cd_shear_lub_final}
\end{equation}
The results show that the leading shear drag decreases inversely with $\hOmega$. This reduction
is a geometric projection effect: at high rotation rates, the surface traction is
directed mainly azimuthally, and only a fraction proportional to
$U/(\Omega R)=\hOmega^{-1}$ contributes to the settling drag.

\subsection{Pressure contribution}
We next estimate the pressure contribution under the assumption that
the unyielded material behaves locally as a
stationary plug or $
u_\theta^y\simeq0,\,\,\,
u_r^y\simeq0
\,\,\,
\text{at}
\,\,\,
y=h(\theta),$ 
 where the superscript $y$ denotes the velocity at the yield surface.
Let $u_\theta(y,\theta)$ be the meridional velocity in the thin layer and
let $p=p(\theta)$ be the leading-order pressure, which is approximately
uniform across the film thickness. The meridional momentum balance is
\begin{equation}
\frac{\partial \tau_{\theta y}}{\partial y}
=
\frac{1}{R}\frac{dp}{d\theta}.
\label{eq:momentum_pressure_lub_final}
\end{equation}
Because the azimuthal shear dominates the local deformation, the meridional
perturbation may be represented using an effective transverse viscosity,
\begin{equation}
\eta_\perp
\sim
k\dot{\gamma}_\Omega^{\,n-1}
+
\frac{\sigma_y}{\dot{\gamma}_\Omega}.
\label{eq:eta_perp_lub_final}
\end{equation}
In the large-$Bi_\Omega$ thin-layer limit, the plastic contribution dominates,
giving
\begin{equation}
\eta_\perp
\sim
\frac{\sigma_y h}{\Omega R\sin\theta}.
\label{eq:eta_perp_plastic_lub_final}
\end{equation}

With $u_\theta(0,\theta)=u_\theta^s=-U\sin\theta$ and
$u_\theta(h,\theta)=0$, the depth-integrated meridional flux is
\begin{equation}
q_\theta
=
\int_0^h u_\theta\,dy
=
\frac{h}{2}u_\theta^s
-
\frac{h^3}{12\eta_\perp R}
\frac{dp}{d\theta}.
\label{eq:flux_stationary_lub_final}
\end{equation}
Mass conservation in the thin layer gives
\begin{equation}
\frac{1}{R\sin\theta}
\frac{d}{d\theta}
\left(
\sin\theta\,q_\theta
\right)
=
u_r^s-u_r^y.
\label{eq:continuity_stationary_lub_final}
\end{equation}
Using $u_r^s=U\cos\theta$ and $u_r^y=0$, substitution of
Eq.~(\ref{eq:flux_stationary_lub_final}) yields the Reynolds-type equation
\begin{equation}
\frac{1}{R^2\sin\theta}
\frac{d}{d\theta}
\left[
\sin\theta
\frac{h^3}{12\eta_\perp}
\frac{dp}{d\theta}
\right]
=
-\frac{U}{2R\sin\theta}
\frac{d}{d\theta}
\left[
h\sin^2\theta
\right]
-
U\cos\theta .
\label{eq:reynolds_stationary_lub_final}
\end{equation}
The first term on the right-hand side represents tangential Couette pumping,
whereas the second term represents normal squeezing due to sedimentation relative
to the stationary plug. In the thin-layer limit, the squeeze term provides the
dominant pressure scale.

The pressure contribution to the axial drag is
\begin{equation}
F_{d,p}
=
2\pi R^2
\int_0^\pi
p(\theta)\cos\theta\sin\theta\,d\theta ,
\label{eq:pressure_drag_lub_final}
\end{equation}
where an arbitrary constant added to $p$ does not contribute to the drag. A
scaling estimate follows from Eq.~(\ref{eq:reynolds_stationary_lub_final}):
\begin{equation}
p
\sim
\eta_\perp U\frac{R^2}{h^3},
\qquad
F_{d,p}
\sim
\eta_\perp U\frac{R^4}{h^3}.
\label{eq:pressure_scale_lub_final}
\end{equation}
Using Eq.~(\ref{eq:eta_perp_plastic_lub_final}) gives
\begin{equation}
F_{d,p}
\sim
\sigma_y R^2
\left(
\frac{U}{\Omega R}
\right)
\left(
\frac{R}{h}
\right)^2 .
\label{eq:Fp_scale_lub_final}
\end{equation}
Explicitly written for a sphere,
\begin{equation}
C_{d,p}^*
=
\frac{F_{d,p}}{\pi R^2\sigma_y}
\sim
C_p\hOmega^{-1}
\left(
\frac{R}{h}
\right)^2,
\label{eq:Cd_pressure_lub_final}
\end{equation}
where $C_p=O(1)$ depends on the detailed pressure distribution. Substituting the
film-thickness estimate in Eq.~(\ref{eq:h_BiOmega_lub_final}) gives, away from
the polar regions,
\begin{equation}
C_{d,p}^*
\sim
C_p
Bi^{2/(n+1)}
\hOmega^{-(3n+1)/(n+1)} .
\label{eq:Cd_pressure_Bi_hOmega_lub_final}
\end{equation}
In the Bingham limit, $n=1$, this reduces to
\begin{equation}
C_{d,p}^*
\sim
C_p\frac{Bi}{\hOmega^2}.
\label{eq:Cd_pressure_bingham_lub_final}
\end{equation}

\subsection{Total drag and gravitational yield number}

Combining the shear and pressure estimates gives the 
lubrication approximation of
\begin{equation}
C_d^*
=
C_{d,\tau}^*
+
C_{d,p}^*
\sim
C_\tau(1+\hOmega^2)^{-1/2}
+
C_p\hOmega^{-1}
\left(
C_\kappa Bi_\Omega
\right)^{2/(n+1)} ,
\label{eq:Cd_total_lub_final}
\end{equation}
where $C_\tau$ and $C_p$ are order-one constants. The first term is the direct
shear contribution and decreases with $\hOmega$ because of the projection of the
surface traction onto the settling direction. The second term is the pressure
contribution generated by lubrication flow through a thin yielded layer bounded
by a stationary outer plug.

The estimate in Eq.~(\ref{eq:Cd_total_lub_final}) should be
interpreted as the large-$Bi$ confined-layer limit. At fixed large $\hOmega$, the
pressure term increases with $Bi_\Omega$ because the yielded layer becomes
thinner and the fluid displaced by sedimentation must pass through a more
confined region. Thus, in the asymptotic stationary-plug limit, increasing $Bi$
is expected to increase the pressure contribution to $C_d^*$. However, the fully
resolved simulations of Fig. \ref{fig:FdversusBi_COMP}c show that, for high-$\hOmega$ 
cases, $C_d^*$ may first
decrease and reach a local minimum at intermediate $Bi$ before increasing again
at larger $Bi$. This behavior indicates a transition in the shape
of the yielded region as discussed in the paper.
 At intermediate $Bi$, rotation can expand the
yielded region around the particle and partially relieve the pressure build-up,
thereby reducing the total drag. This  effect may be represented phenomenologically by replacing  $C_p$ with $C_{p0}\, \mathcal{M}(Bi,\hOmega)$ where $\mathcal{M}<1$ represents the increased mobility of the connected yielded region. 

At larger $Bi$, the yielded region contracts
toward the particle surface, the locally confined-layer assumption becomes more
appropriate, and the pressure contribution increases. The observed local minimum
in $C_d^*$ therefore marks the transition from a rotation-assisted,
pressure-relieved regime to a high-$Bi$ confined lubrication regime.

In either cases, the relative importance of the two terms is obtained from
Eqs.~(\ref{eq:Cd_shear_lub_final}) and
(\ref{eq:Cd_pressure_Bi_hOmega_lub_final}):
\begin{equation}
\frac{C_{d,p}^*}{C_{d,\tau}^*}
\sim
C_{p\tau}
\left(
C_\kappa Bi_\Omega
\right)^{2/(n+1)} .
\label{eq:pressure_shear_ratio_lub_final}
\end{equation}
This estimate explains why pressure drag can become large when the yield surface
is close to the particle. 
We can now estimate the gravitational yield number for a sedimenting rotating sphere,
\begin{equation}
Y_G(Bi,\hOmega)
\sim
\frac{2/3}
{
C_\tau(1+\hOmega^2)^{-1/2}
+
C_p\hOmega^{-1}
\left(
C_\kappa Bi_\Omega
\right)^{2/(n+1)}
}.
\label{eq:YG_lub_final}
\end{equation}
This expression provides a compact scaling connection between rotation, the
Bingham number, and the apparent gravitational yield number. Because
$Y_G=(2/3)/C_d^*$ for a sphere, any local minimum in $C_d^*$ corresponds to a
local maximum in $Y_G$. Therefore, the local minimum in $C_d^*$ observed in the
fully resolved high-$\hOmega$ simulations implies that the apparent yield number
can increase at intermediate $Bi$ and then decrease again as $Bi$ becomes large.
In the shear-dominated high-swirl limit, the pressure term is negligible and
\begin{equation}
Y_G
\sim
\frac{2}{3 C_\tau}
(1+\hOmega^2)^{1/2}
\sim
\frac{2}{3 C_\tau}\hOmega,
\qquad
\hOmega\gg1 .
\label{eq:YG_shear_lub_final}
\end{equation}
In contrast, if the pressure term dominates, then for a Bingham material
\begin{equation}
Y_G
\sim
\frac{2}{3\,C_p}
\frac{\hOmega^2}{Bi}.
\label{eq:YG_pressure_bingham_lub_final}
\end{equation}
Thus, the variation of $Y_G$ with $Bi$ at fixed high $\hOmega$ need not be
monotonic. At intermediate $Bi$, rotation-assisted yielding can reduce
$C_d^*$ and increase $Y_G$, whereas at sufficiently large $Bi$ the yielded layer
becomes thin and confined, pressure drag increases, and $Y_G$ decreases. This
interpretation is consistent with the local minimum in $C_d^*$ observed in the
fully resolved simulations (Fig.~\ref{fig:FdversusBi_COMP}c).
Thus, the apparent gravitational yield number increases with rotation because
rotation reduces the drag required to maintain a given settling velocity.
However, the rate of this increase depends on whether the resistance is governed
primarily by shear projection or by pressure build-up in the confined yielded
layer.

\subsection{Range of validity}
These scaling relations are derived based on four assumptions. \\
(1) The flow must be rotation-dominated:
\begin{equation}
\hOmega=\frac{R\Omega}{U}\gg1 .
\label{eq:validity_hOmega_lub_final}
\end{equation}
(2) The yielded layer must be thin and close to the interface:
\begin{equation}
\frac{h}{R}\ll1,
\qquad
\frac{h}{R}
\sim
\left[
\frac{\sin^n\theta}
{C_\kappa Bi_\Omega}
\right]^{1/(n+1)} ,
\label{eq:validity_h_lub_final}
\end{equation}
which requires $Bi_\Omega\gg1$ away from the polar regions or equivalently  $\hOmega^n\ll Bi$.
\\
(3) The rotation-controlled estimate of $h$ assumes that the pressure stress
does not reorganize the yield surface. A useful consistency requirement is
\begin{equation}
\hOmega
\gg
\left(
C_\kappa Bi_\Omega
\right)^{2/(n+1)} .
\label{eq:validity_pressure_lub_final}
\end{equation}
Combining this with $Bi_\Omega\gg1$ gives
\begin{equation}
{Bi}^{\frac{2}{3n+1}}\ll \hOmega \ll Bi^{\frac{1}{n}} .
\label{eq:validity_window_bingham_lub_final}
\end{equation}
which for Bingham flow translates to
\begin{equation}
\sqrt{Bi}\ll \hOmega \ll Bi .
\label{eq:validity_window_bingham_lub_final}
\end{equation}
Within this window, the yielded layer is thin, its thickness is primarily set by
rotational shear, and the stationary-plug pressure estimate is asymptotically
consistent.
\\
(4) The analysis is not valid near $\theta=0$ and $\theta=\pi$, where
$\sin\theta\rightarrow0$ and the rotational surface velocity vanishes. Those
polar regions require a separate local description.

For scaling analysis, inside these small polar regions, the
local flow can be approximated  as squeeze flow between the particle surface and the
locally stationary outer plug. Introducing the local polar coordinate
$\rho=R\theta$ near $\theta=0$ and
$\rho=R(\pi-\theta)$ near $\theta=\pi$, the normal velocity remains
$O(U)$, while the rotational velocity scales as $\Omega\rho$. If the polar cap
has radius $\ell_p$ and representative gap thickness $h_p$, the local
Reynolds-type balance gives
\begin{equation}
F_{d,p}^{\mathrm{polar}}
\sim
\eta_{\mathrm{eff},p}
U\frac{\ell_p^4}{h_p^3}.
\label{eq:polar_pressure_force_compact}
\end{equation}

The effective viscosity is evaluated using the squeeze-induced shear rate,
\begin{equation}
\dot{\gamma}_p
\sim
\frac{U\ell_p}{h_p^2},
\qquad
\eta_{\mathrm{eff},p}
\sim
k\dot{\gamma}_p^{\,n-1}
+
\frac{\sigma_y}{\dot{\gamma}_p}.
\label{eq:polar_effective_viscosity_compact}
\end{equation}
When the yield-stress contribution dominates, this reduces to
\begin{equation}
F_{d,p}^{\mathrm{polar}}
\sim
\sigma_y\frac{\ell_p^3}{h_p},
\qquad
C_{d,p}^{*,\mathrm{polar}}
\sim
C_{p0}
\left(\frac{\ell_p}{R}\right)^2
\left(\frac{\ell_p}{h_p}\right),
\label{eq:polar_Cd_compact}
\end{equation}
where $C_{p0}=O(1)$. The polar cap size follows by matching the squeeze-induced
shear to the outer rotational shear,
\begin{equation}
\frac{\Omega\ell_p}{h_p}
\sim
\frac{U\ell_p}{h_p^2},
\qquad
\Rightarrow
\qquad
\frac{h_p}{R}\sim\hOmega^{-1}.
\label{eq:polar_matching_compact}
\end{equation}

Using the outer estimate
$h/R\sim[\theta^n/(C_\kappa Bi_\Omega)]^{1/(n+1)}$ gives
\begin{equation}
\theta_c=\frac{\ell_p}{R}
\sim
\left(C_\kappa Bi\right)^{1/n}
\hOmega^{-(2+1/n)},
\label{eq:polar_cap_compact}
\end{equation}
and, for a Bingham material results in $
\theta_c\sim C_\kappa Bi\,\hOmega^{-3}
$. This means that 
when $\theta_c\ll1$, the polar caps mainly regularize the outer
solution and modify the order-one pressure coefficient. If $\theta_c$ is not
small, the pressure field and yielded-layer topology must be obtained from the
full viscoplastic solution.

}

\bibliographystyle{jfm}
\bibliography{jfm}



\end{document}